\documentclass[twocolumn,aps,floatfix,superscriptaddress]{revtex4}
\pdfoutput=1 
\usepackage{amsmath,amssymb,eucal,graphicx,float,epstopdf}

\def\beq{\begin{eqnarray}}
\def\eeq{\end{eqnarray}}
\def\be{\begin{equation}}
\def\ee{\end{equation}}
\def\eq{&=&}

\def\lb{\label}
\def\q{\quad}
\def\qq{\qquad}

\newcommand \nt   {\nonumber \\ }

\begin{document}
\bibliographystyle{apsrev}

\title{Phase Transitions in Systems with Aggregation and Shattering }
\author{P. L. Krapivsky}
\affiliation{Department of Physics, Boston University, Boston, MA 02215, USA}
\author{W. Otieno}
\affiliation{Department of Mathematics, University of Leicester, Leicester LE1 7RH, UK}
\author{N. V. Brilliantov}
\affiliation{Department of Mathematics, University of Leicester, Leicester LE1 7RH, UK}

\begin{abstract}
We consider a system of clusters made of elementary building blocks, monomers, and evolving via collisions between diffusing monomers and immobile composite clusters. In our model, the cluster-monomer collision can lead to the attachment of the monomer to the cluster (addition process) or to the total break-up of the cluster (shattering process). A phase transition, separating qualitatively different behaviors, occurs when the probability of shattering events exceeds a certain threshold. The novel feature of the phase transition is the dramatic dependence on the initial conditions. 
\end{abstract}

\maketitle

\section{Introduction}

Addition is the basic growth mechanism generating objects of potentially unlimited size.  In the simplest implementation the process begins with a vast number of identical elementary building blocks, monomers, and larger objects are formed by adding monomers. The smallest composite objects, dimers, arise via the reaction process $ M+M\to I_2$, where $M =I_1$ denotes a monomer. A trimer is formed by adding a monomer to a dimer, $M+I_2\to I_3$,  and generally
\begin{equation}
\label{scheme} 
M + I_k \mathop{\longrightarrow}^{A_k}  I_{k+1}
\end{equation}
We tacitly assume that each cluster is fully described by a single parameter,  its mass $k$ which is the
number of constituent monomers; the addition process \eqref{scheme} is then characterized by the collection
of addition rates $A_k$.

The addition process \eqref{scheme} mimics aggregating systems in which monomers are mobile,  while composite objects (clusters) are  immovable. In contrast to the classical aggregation kinetics where all clusters react with each other \cite{smoluchowski1916drei,Krapivsky,Leyvraz}, clusters do not directly interact in the process \eqref{scheme}, they just gradually grow by the addition of monomers. The model \eqref{scheme} has been used to mimic numerous processes, see e.g. \cite{bk,Blackman91,Blackman94,Privman1999,Laurencot1999,Privman2010,Privman2013} and references therein. Addition processes also underlie self-assembly, see \cite{RW04,SA08,Privman2009,Igor14,Erik17}. One important application of this model is to surface science where the monomers are adatoms hopping on the substrate, see e.g. \cite{Evans_PRB,Wolf,MBE,Zinke_Allmang1999,Krapivsky_PRB1999,Family2001,Family_PRB2001}. When two adatoms meet they form an {\em immobile} island, a dimer; similarly when an adatom meets an island $I_k$, it attaches irreversibly
forming an island $I_{k+1}$ of mass $k+1$.

The opposite processes of clusters decomposition are also possible when a cluster collides with an energetic
monomer. In this case the energy of the adatom, transmitted to the island, can break some bonds between adatoms comprising the island. Here we investigate the extreme version positing that clusters undergo a total break-up (shattering) into constituting monomers:
\begin{equation}
\label{scheme:shattering} 
M+I_{k}\mathop{\longrightarrow}^{S_k} \underbrace{M+\ldots+M}_{k+1},
\end{equation}
where $k\geq 2$ and $S_k$ are the shattering rates. As we show below, the interplay between addition and
shattering generically results in a phase transition. Namely, with increasing shattering rates the stationary
concentrations of clusters  undergo a discontinuous jump when shattering rates exceeds some
threshold. In particular, if the shattering events are rare the stationary concentration of monomers
vanishes, while it becomes finite above the critical  shattering rate. This critical rate demarcates
qualitatively different states of the system: For the super-critical shattering the system reaches an equilibrium
stationary state independent on the initial conditions. For the sub-critical
shattering, the stationary states are jammed, that is, they are determined by the initial conditions.
Surprisingly, the critical shattering depends on the initial conditions of the system itself.

\section{Addition and Shattering with Mass-Independent Rates}
\label{sec:const}

We start our analysis with the simplest model in which the addition and shattering rates are mass independent:
\begin{equation}
\label{AS:const}
A_k = 1, \qquad S_k = \lambda
\end{equation}
In the context of surface science, the mass-independent addition rate is particularly natural in the realm of the point-island model (where each island occupies a single lattice site), and in the most interesting case of two-dimensional substrate it provides a good approximation in more realistic cases \cite{Evans_PRB,Wolf,MBE,Zinke_Allmang1999,Krapivsky_PRB1999}. 

The evolution equations for the addition-and-shattering model with rates \eqref{AS:const} read
\begin{subequations}
\begin{align}
&\frac{d c_1}{dt}=-c_1^2 - c_1 \sum_{j\geq 1} c_j  + \lambda c_1 \sum_{j\geq 2} jc_j, 
\label{c1}\\
&\frac{d c_k}{dt}=c_1c_{k-1}-c_1c_k - \lambda c_1 c_k, \quad k\geq 2.
\label{ck}
\end{align}
\end{subequations}
Here $c_k$ is the density of clusters of mass $k$, so $k=1$ corresponds to mobile adatoms and $k\geq 2$
describe immobile islands. These equations are the straightforward generalization of the addition model \cite{bk} where $\lambda=0$. The first and second terms in the right-hand side of Eq.~\eqref{c1} give the rate of monomers loss due to aggregation, while the third term quantifies the gain of monomers in the shattering events \eqref{scheme:shattering} with $S_k =\lambda$. Similarly, the three terms in the right-hand side  of Eq.~\eqref{ck} describe respectively the gain of $k$-mers in the reactions of monomers with the $(k-1)$-mers and loss of the $k$-mers in the aggregation and shattering
processes. 

Using \eqref{c1}--\eqref{ck} one can verify that the mass density $\sum_{j\geq 1} jc_j$ remains constant throughout the evolution. We shall set the mass density to unity if not stated otherwise
\begin{equation}
\label{mass}
M=\sum_{j\geq 1} jc_j=1
\end{equation}
if not stated otherwise. This can be done without loss of generality.  Indeed,  the right-hand side of the rate equations \eqref{c1}--\eqref{ck} are quadratic polynomials and hence the rate equations are invariant under the transformation $t\to t/M$ and $c_k\to M c_k$. With this transformation one can always set the mass density to unity. 

Introducing the auxiliary time
\begin{equation}
\label{time} \tau=\int_0^t dt'\, c_1(t')
\end{equation}
we linearize the above equations
\begin{subequations}
\begin{align}
&\frac{d c_1}{d\tau}= \lambda - (1+\lambda)c_1 - N, \qq N=\sum_{j\geq 1} c_j
\label{1AS}\\
&\frac{d c_k}{d\tau}=c_{k-1}-(1+\lambda)c_k, \qq ~~k\geq 2
\label{2AS}\\
&\frac{d N}{d\tau}=\lambda - (1+\lambda)N. \label{3AS}
\end{align}
\end{subequations}
We used the relation $\sum_{j\geq 2} jc_j=1-c_1$ which follows from \eqref{mass} and displayed the
rate equation for the total cluster density, $N(\tau)$, obtained by summing \eqref{1AS} and Eqs.~\eqref{2AS} for
all $k\geq 2$.

We shall always use the mono-disperse initial condition
\begin{equation}
\label{IC:mono}
c_k(0) = \delta_{k,1}
\end{equation}
if not stated otherwise. Solving \eqref{3AS} subject to $N(0)=1$ gives
\begin{equation}
\label{NAS} N= \frac{\lambda + e^{-(1+\lambda)\tau}}{1+\lambda}\,.
\end{equation}
Plugging \eqref{NAS} into \eqref{1AS} we obtain a close equation for the density of monomers which is solved
to yield
\begin{equation}
\label{c1AS} c_1= \left[1-\tfrac{\tau}{1+\lambda}\right] e^{-(1+\lambda)\tau} +
\tfrac{\lambda^2}{(1+\lambda)^2} \left[1 -  e^{-(1+\lambda)\tau}\right].
\end{equation}
To find the evolution of the island densities, we apply to Eqs.~\eqref{2AS} the Laplace transform,
$\widehat{c}_k =\int_0^{\infty} c_k(\tau)e^{-p \tau} d\tau$. Since $c_k(0)=0$ for $k\geq 2$, we obtain

\be 
\lb{eq:Laplace} 
p\widehat{c}_k = \widehat{c}_{k-1} -(1+\lambda) \widehat{c}_k,  \qq k \geq 2, \ee
from which 

\be \lb{eq:Laplace_c1ck} \widehat{c}_k(p) = \frac{\widehat{c}_{1}(p)}{(p+1+\lambda)^{k-1}},   \qq k \geq 2. \ee
It is straightforward to find the Laplace transform of $c_1(\tau)$ given by \eqref{c1AS}; substituting the
result into Eq.~\eqref{eq:Laplace_c1ck} and performing the inverse Laplace transform we obtain 

\begin{eqnarray}
\label{ck:sol} 
c_{k+1}(\tau) &=& \frac{\tau^k}{\Lambda k!} \left[ \frac{2 \Lambda-1}{\Lambda} -
\frac{\tau}{k+1} \right] e^{-\Lambda \tau} \nonumber\\
&+& \frac{(\Lambda-1)^2  \gamma\left( k, \Lambda \tau \right)
}{\Lambda^{k+2} (k-1)!}
\end{eqnarray}
where
\begin{equation}
\label{G:incomplete} \gamma(k,a)= \int_0^a du\,u^{k-1} e^{-u}
\end{equation}
is the incomplete Gamma function. Hereinafter we often use the shorthand notation $\Lambda \equiv \lambda+1$. If we assume that $\tau \to \infty$ as $t \to \infty$, Eqs.~\eqref{NAS} and \eqref{ck:sol} give the asymptotic distribution for clusters
size and total cluster density:

\beq \lb{cksimple} C_k  \equiv c_k(\tau=\infty)  \eq  \frac{\lambda^2}{(1+\lambda)^{k+1}}  \qq k\geq 1 ,  \\
\lb{Nsimple}N(\tau = \infty) \eq  \frac{\lambda}{1+\lambda}  \,. \eeq
Exactly the same result is obtained if one seeks the stationary solution to Eqs.~\eqref{1AS}--\eqref{3AS}.

\begin{figure}
\centering
\includegraphics[width=8.5cm]{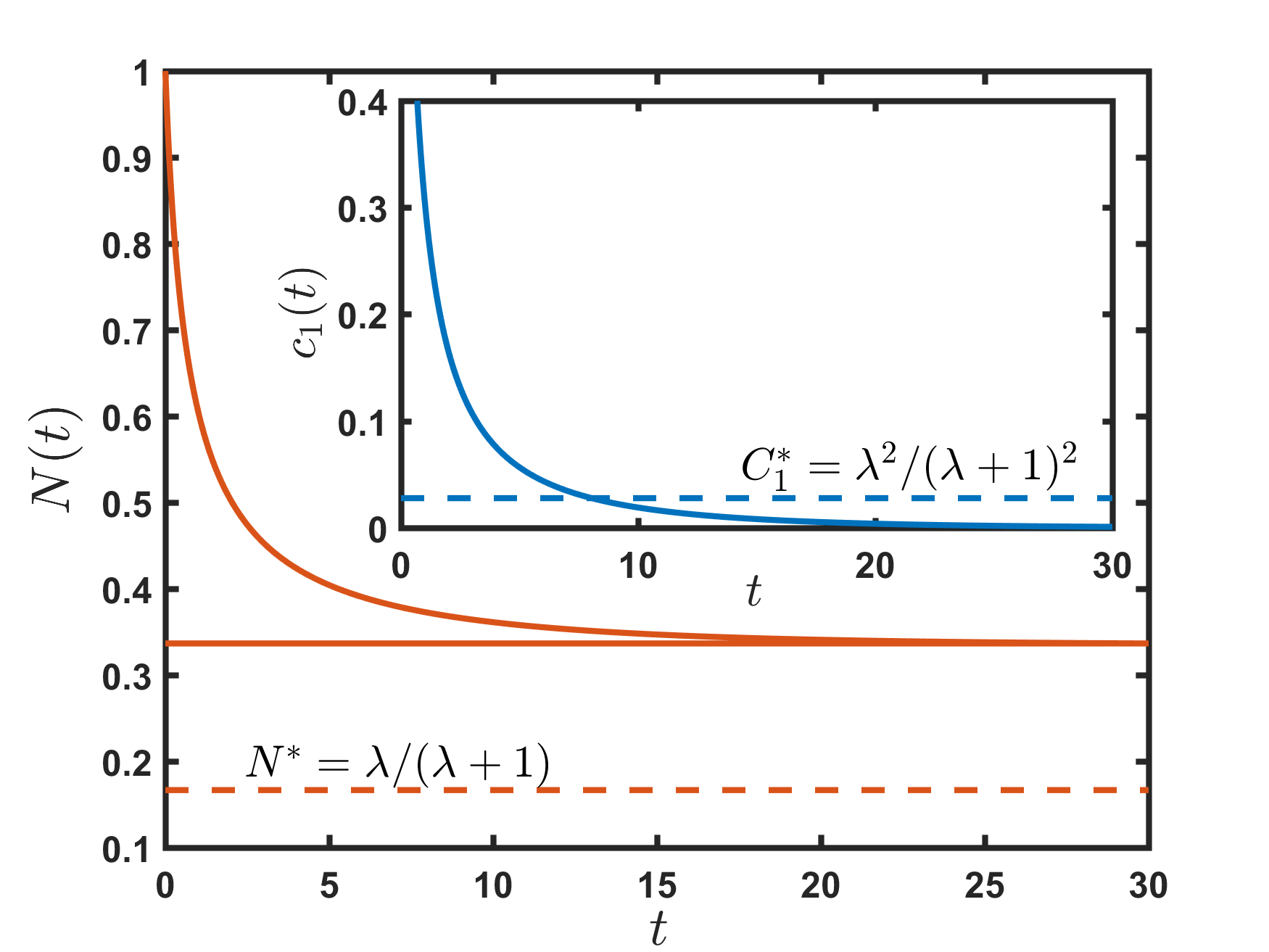}
\includegraphics[width=8.5cm]{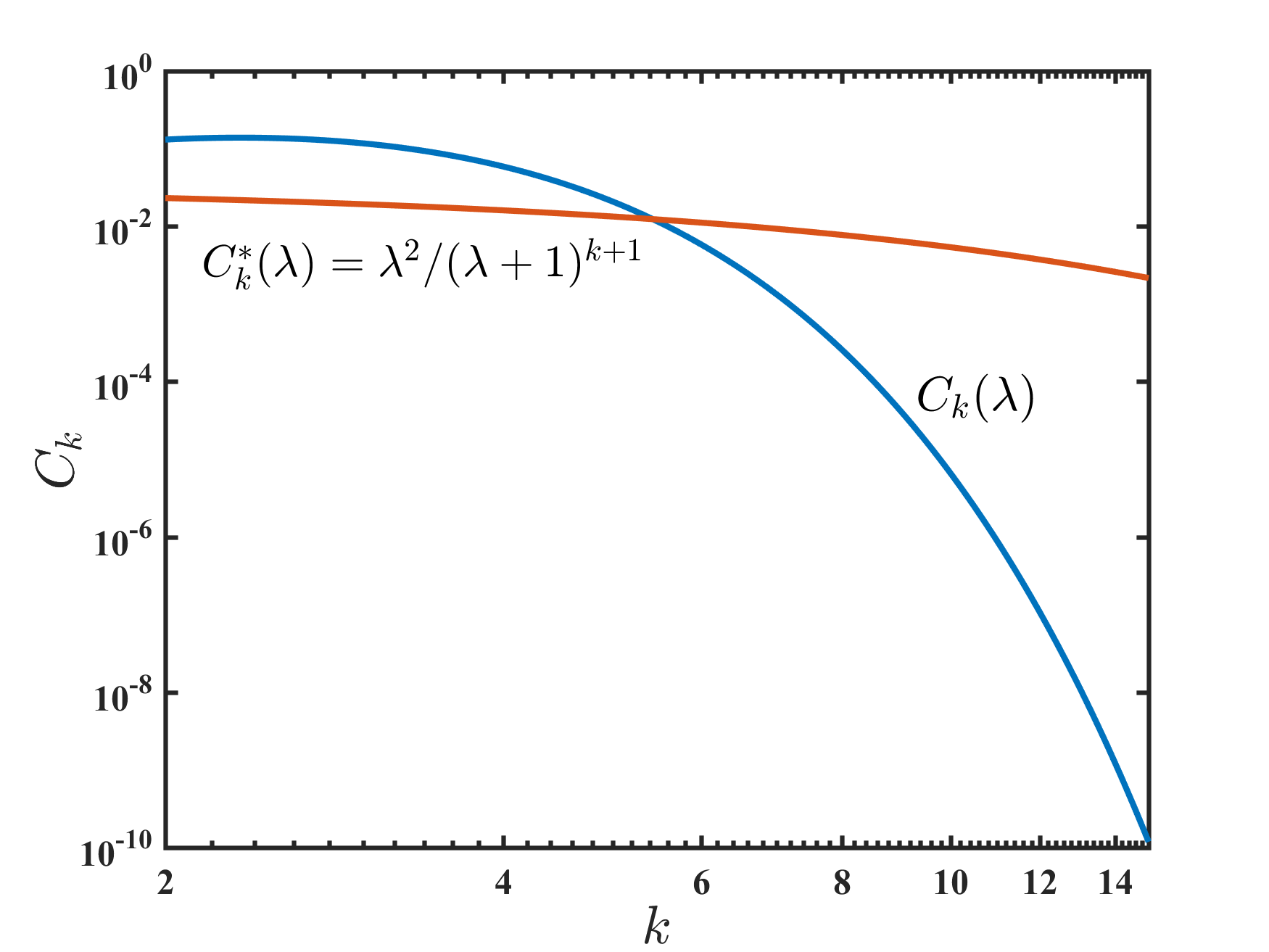}
\caption{Top: Evolution of the monomer density $c_1(t)$ and total cluster density $N(t)$ for the case of constant kernels and the mono-disperse initial condition when $\lambda =0.2$. The system approaches a final state $c_1(t=\infty)$ and $N(t=\infty)$ which differs from the equilibrium state \eqref{cksimple}--\eqref{Nsimple}. Bottom: Asymptotic
cluster size distribution, $c_k(t=\infty)$, dramatically differs from the equilibrium distribution~\eqref{cksimple}. }\label{fig:c1Nsubc}
\end{figure}

Naively, one would expect that the system evolves to the equilibrium state \eqref{cksimple}--\eqref{Nsimple} independently on initial conditions. Yet the system demonstrates a richer behavior, see  Fig.~\ref{fig:c1Nsubc}. As one can see from Fig.~\ref{fig:c1Nsubc}, the densities m ay relax to a state dramatically different from the equilibrium solution  \eqref{cksimple}--\eqref{Nsimple}. 

\begin{figure}
\centering
\includegraphics[width=8.5cm]{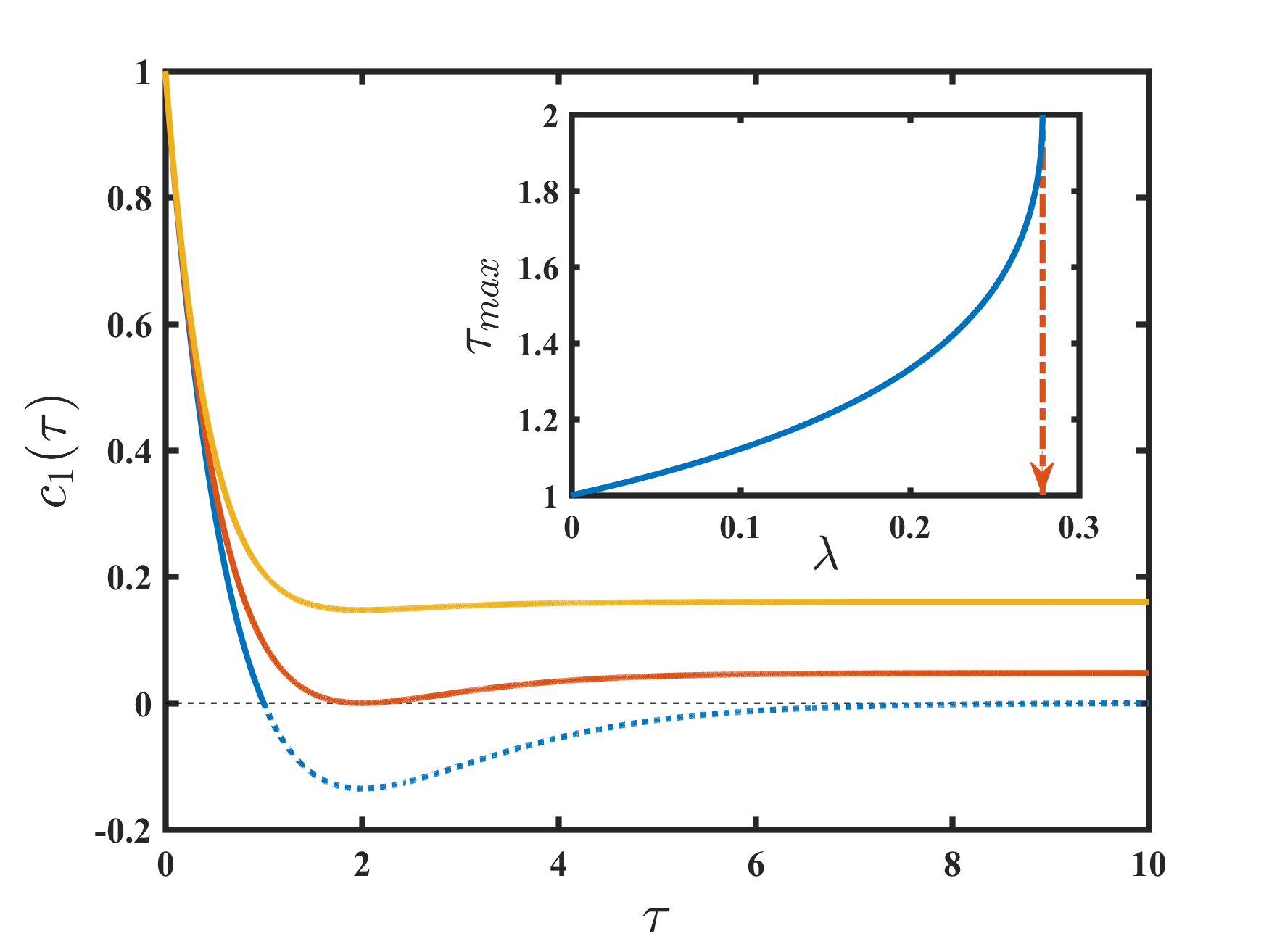}
\caption{The monomer density $c_1(\tau)$ remains positive when the shattering rate satisfies
$\lambda>\lambda_c\approx 0.27846$. Shown (bottom to top) is the monomer density for $\lambda=0, \lambda_c,
2/3$. Inset: When physical time diverges, $t\to \infty$, the auxiliary time for $\lambda<\lambda_c$,
remains finite, $\tau \to \tau_{\rm max} (\lambda)$. } 
\label{Fig:monAS}
\end{figure}

To understand this surprising behavior we notice that the monomer density given by \eqref{c1AS} is physically
applicable as long as $c_1\geq 0$. For sufficiently large shattering rates the monomer density is always
positive, while for smaller rates $c_1(\tau)$ first vanishes at a certain $\tau_\text{max}(\lambda) $. In
these situations, the physical range $0\leq t<\infty$ corresponds to $0\leq \tau<\tau_\text{max}(\lambda)$,
so that the above assumption, that $t \to \infty$ as $\tau \to \infty$ is not valid.  More precisely,
$\tau_\text{max}(\lambda)$ is finite when $\lambda\leq \lambda_c$, while for $\lambda> \lambda_c$ the monomer
density remains positive for all $0<\tau<\infty$. This is obvious from Fig.~\ref{Fig:monAS}. To establish
this assertion analytically we first notice that $c_1(\tau)$ given by \eqref{c1AS} reaches minimum when
$\tau=2$, from which the minimal possible value is
\begin{equation*}
c_1^{\rm{min}}=\left[1-\frac{2}{1+\lambda}\right] e^{-2(1+\lambda)} + \frac{\lambda^2}{(1+\lambda)^2} \left[1
- e^{-2(1+\lambda)}\right].
\end{equation*}
Analyzing this expression we find that it remains positive when $\lambda>\lambda_c$ where $\lambda_c$ is the
root of $ e^{1+\lambda_c}=1/\lambda_c$, that is,  $\lambda_c = W(1/e)$, where $W(x)$ is the Lambert function;
numerically $\lambda_c = 0.27846\ldots$. 

Different behaviors emerge depending on whether $\lambda$ is smaller, equal, or larger than $\lambda_c$.

\subsection{Sub-critical regime: $\lambda <\lambda_c$} 

In the region $ 0\leq \lambda <\lambda_c$ the monomer density vanishes at a certain $\tau_\text{max}(\lambda)$ which is found from \eqref{c1AS} to be the root of the transcendental equation
\begin{equation}
\label{tau:max} 
e^{(1+\lambda)\tau_\text{max}} = \frac{(1+\lambda)\tau_\text{max}-1-2\lambda}{\lambda^2}
\end{equation}
The solution of Eq.~\eqref{tau:max} may be expressed in terms of the Lambert function:
\begin{equation}
\label{tau:max_Lam} 
\tau_\text{max} = \frac{1+2\lambda-W\left( -\lambda^2 e^{2\lambda +1} \right) }{
1+\lambda}
\end{equation}
Note that $\tau_\text{max}$ increases from 1 to 2 as $\lambda$ increases from 0 to $\lambda_c$, see
Fig.~\ref{Fig:monAS}. For $\tau_\text{max}=1$ and $\lambda =0$, Eq.~\eqref{ck:sol} reproduces the final
distribution of cluster sizes for the additional aggregation without shattering~\cite{bk}.

The monomer density vanishes exponentially in terms of the physical time,

$$
c_1 \sim e^{-t(N_\infty-\lambda)} \qq t \to \infty,
$$
and the island densities saturate at $t\to \infty$, that is,  $C_k(\lambda) \equiv
c_k(\tau_\text{max})>0$ for $k\geq 2$, with $c_k(\tau)$ and $\tau_\text{max}$ given respectively by
Eqs.~\eqref{ck:sol} and \eqref{tau:max_Lam}; the same is true for the final density of islands
$N_\infty(\lambda)=N[\tau_\text{max}(\lambda)]$ which is  positive for all $\lambda\geq 0$.

\begin{figure}
\centering
\includegraphics[width=8.5cm]{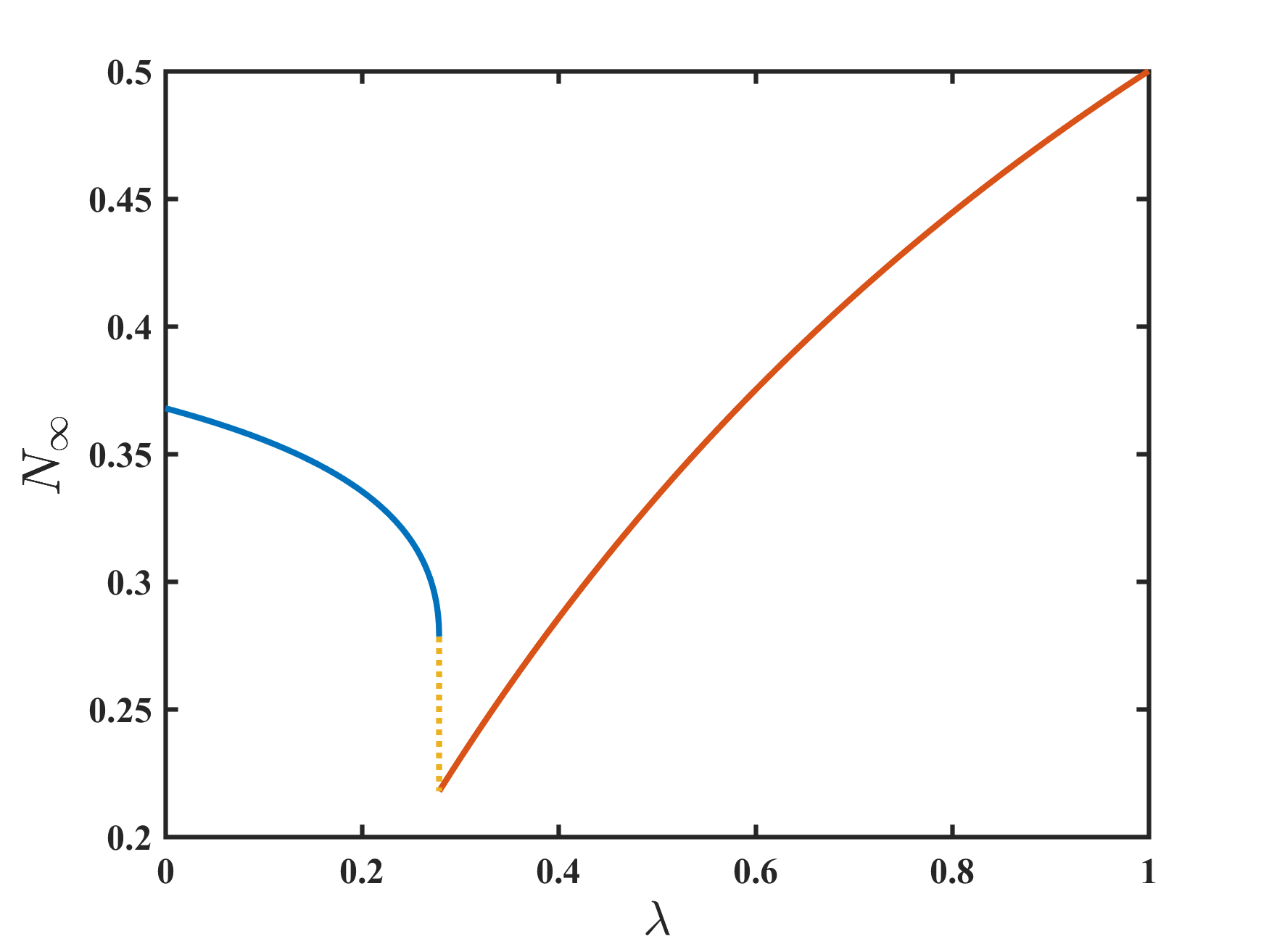}
\caption{The final density of clusters $N_\infty(\lambda)$ is a  decreasing function of $\lambda$ in the
range $0\leq \lambda\leq \lambda_c\approx 0.27846$, it undergoes a first order transition at $\lambda_c$, and
becomes an increasing function of $\lambda$ in the range $\lambda>\lambda_c$ where
$N_\infty=N^*=\lambda/(1+\lambda)$. } \label{Fig:N_inf}
\end{figure}

Analyzing \eqref{NAS} and \eqref{tau:max} one finds that in the subcritical region $N_\infty(\lambda)$ is a
decreasing function of the shattering rate $\lambda$, namely it decreases from $N_\infty(0)=e^{-1}\approx
0.36788$ to $N_\infty(\lambda_c)=\lambda_c$, see Fig.~\ref{Fig:N_inf}. The approach of $N_\infty(\lambda)$ to
the density $N_\infty(\lambda_c)=\lambda_c$ in the critical regime is singular:
\begin{equation}
N_\infty(\lambda)-N_\infty(\lambda_c)\simeq 2\lambda_c\sqrt{\frac{\lambda_c}{1+\lambda_c}}\,\sqrt{\lambda_c-\lambda}
\end{equation}
as $\lambda\uparrow \lambda_c$.

\subsection{Critical regime: $\lambda = \lambda_c$}
\label{subsec:crit}

In the critical regime the monomer density \eqref{c1AS} also vanishes. More precisely, when $\tau\to \tau_\text{max}(\lambda_c)=2$, the monomer density decreases as $c_1 \simeq \tfrac{1}{2}\lambda_c^2\,(2-\tau)^2$ which can be re-written as
\begin{equation}
\label{mon:crit}
c_1(t) \simeq \frac{2}{\lambda_c^2}\,\frac{1}{t^2}
\end{equation}
in terms of the original time variable. Thus in the critical regime the process slows down --- the vanishing is algebraic rather than exponential.

The total density of clusters also exhibits an algebraic approach to the final density:
\begin{equation}
\label{total:crit}
N(t) - N_\infty \simeq \frac{2}{t}
\end{equation}
where $N_\infty = N_\infty(\lambda_c) = \lambda_c$. The  asymptotic \eqref{total:crit} follows from
\eqref{NAS}. The decay exponent in \eqref{total:crit} is twice smaller than the exponent characterizing the
decay of monomers.

The approach to the final density for any species of islands is  similar to \eqref{total:crit}, viz.
\begin{equation}
c_k(t) - C_k \simeq \frac{B_k}{t}.
\end{equation}
Using \eqref{ck} one can express the amplitudes $B_k$ through the  final densities $C_k\equiv c_k(\tau=2)$:
\begin{equation}
\label{BC} B_k = 2\,\frac{\Lambda_c C_k - C_{k-1} }{(\Lambda_c - 1)^2}\,, \qquad \Lambda_c=1+\lambda_c
\end{equation}
for $k\geq 2$. The final densities can be expressed using Eq.~\eqref{ck:sol}  with $\tau_{\rm max} =2$. One finds
\begin{eqnarray}
\label{Ck:crit}
\frac{C_{k+1}}{(\Lambda_c-1)^2} &=& \frac{2^k}{\Lambda_c k!} \left[ \frac{2
\Lambda_c-1}{\Lambda_c} - \frac{2}{k+1} \right]  \nonumber\\
&+& \frac{ \gamma\left( k, 2\Lambda_c \right) }{\Lambda_c^{k+2} (k-1)!}
\end{eqnarray}
Combining \eqref{Ck:crit} with \eqref{BC} we get

\be 
\lb{Bk} 
B_k=-2^{k-1}\frac{(k-1)(k-4)}{k!} 
\ee
where we have used the identity
\begin{equation}
\label{G:identity} (k-1)\gamma(k-1,2\Lambda)-\gamma(k,2\Lambda) = (2\Lambda)^{k-1} e^{-2\Lambda}
\end{equation}
which can be derived by substituting the definition \eqref{G:incomplete} into \eqref{G:identity} and
performing the integration by part of the second integral on the left-hand side. We also use the identity
$e^{-\Lambda_c} = \Lambda_c-1$ obeyed by $\Lambda_c=1+\lambda_c$.
Remarkably, the amplitudes $B_k$ are independent on $\Lambda_c$ and rational.

\subsection{Super-critical region:  $\lambda > \lambda_c$}

In the super-critical case $\tau \to \infty$ as $t \to \infty$ and we can use Eq.~\eqref{cksimple} for the
final cluster densities $C_k(\lambda)$:
\begin{equation}
\label{final:super} 
C_k(\lambda) = \frac{\lambda^2}{(1+\lambda)^{k+1}} =C_k^* \qq \q k \geq 1.
\end{equation}

\begin{figure}
\centering
\includegraphics[width=8.5cm]{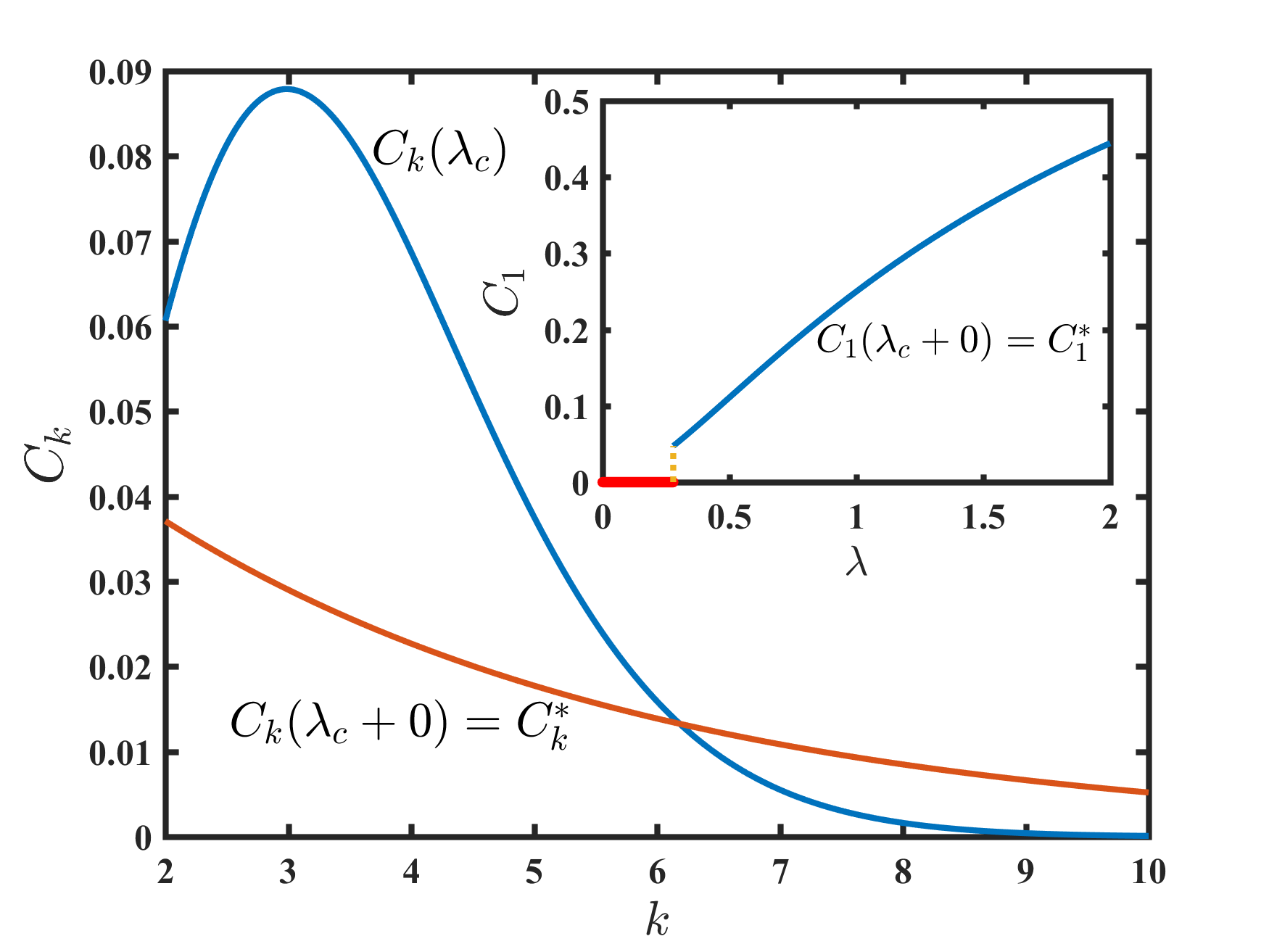} 
\caption{The discontinuous jump in the cluster size distribution from $C_k(\lambda_c)$ to
$C_k(\lambda_c+0)$, given by Eq.~\eqref{cksimple} at the critical point $\lambda=\lambda_c$.
Inset: The final density of monomers $C_1(\lambda)$ vanishes in the range $0\leq \lambda\leq
\lambda_c\approx 0.278465$, undergoes a first order transition at $\lambda_c$, and becomes an increasing
function of $\lambda$ in the super-critical region $\lambda>\lambda_c$ where $C_1=\lambda^2/(1+\lambda)^2$.}
 \label{Fig:C1_inf}
\end{figure}

Thus the final monomer density exhibits a discontinuous (first order) phase transition as a function of
$\lambda$:
\begin{equation}
C_1(\lambda) =
\begin{cases}
0 & \lambda\leq \lambda_c\\
\frac{\lambda^2}{(1+\lambda)^{2}} & \lambda > \lambda_c
\end{cases}
\end{equation}
see Fig.~\ref{Fig:C1_inf}. The total cluster density also exhibits the discontinuous phase transition at
$\lambda= \lambda_c$, more precisely
\begin{equation}
N_{\infty}(\lambda) =
\begin{cases}
\lambda_c  & \lambda = \lambda_c\\
\frac{\lambda_c}{1+\lambda_c } & \lambda = \lambda_c+0
\end{cases}
\end{equation}
The  same is valid for the final densities $C_k$ which undergo a jump at the critical point
$\lambda=\lambda_c$ from the values $C_k(\lambda_c)$, given by Eq.~\eqref{Ck:crit}, to
$C_k(\lambda_c+0)=\lambda_c^2/(1+\lambda_c)^{k+1}$, as it illustrated in Fig.~\ref{Fig:C1_inf}.

\subsection{Dependence on initial conditions}

The previous analysis has been done for the mono-disperse initial condition \eqref{IC:mono}. We now briefly discuss more general initial conditions. (We still set $M=1$.)

The behavior in the super-critical region is universal, e.g. the final  densities are given by
\eqref{final:super} do not depend on the initial condition. The critical shattering rate is not 
universal, however, namely it depends on the initial condition. As an example,
consider the initial condition
\begin{equation}
\label{IC:MD} c_1(0)=c_2(0)=1/3 \,, \quad c_k(0)=0 \quad\text{when}\quad k\geq 3,
\end{equation}
which corresponds to $N(0)=2/3$. Performing the same steps that led to
Eq.~\eqref{c1AS}, we obtain the monomer density for the above initial conditions:
\begin{equation*}
c_1= \tfrac{1}{3}\left[1-\tfrac{2-\lambda}{1+\lambda}\tau\right] e^{-(1+\lambda)\tau}
+ \tfrac{\lambda^2}{(1+\lambda)^2} \left[1 -  e^{-(1+\lambda)\tau}\right].
\end{equation*}
The qualitative behaviors remain the same. Chief quantitative results are also universal, e.g., in the
critical regime the monomer density  exhibits the $t^{-2}$ decay. The critical shattering rate is however
 $\lambda_c\approx 0.30057$ for the initial condition \eqref{IC:MD}.

For an arbitrary initial conditions solution of Eqs.~\eqref{1AS}, \eqref{3AS} yield for the total
cluster and monomer density
\beq
\label{N:gen} N(\tau) &=& \frac{\lambda}{1+\lambda}\left[1 -  e^{-(1+\lambda)\tau}\right]+ N_0\,
e^{-(1+\lambda)\tau} , \\
\label{c1:gen}
c_1(\tau)
& =& c_{1,0} \,e^{-(1+\lambda)\tau}-\left[N_0-\frac{\lambda}{1+\lambda}\right]\tau\, e^{-(1+\lambda)\tau} \nonumber\\
&+& \frac{\lambda^2}{(1+\lambda)^2} \left[1 -  e^{-(1+\lambda)\tau}\right],
\end{eqnarray}
where $N_0=N(0)$ and $c_{1,0}=c_1(0)$.  Preforming again the Laplace transform and using
the Laplace transform of $c_1(\tau)$ given by \eqref{c1:gen}, we find the  clusters size distribution for
arbitrary initial conditions:

\beq 
\label{ckgen:sol} 
c_{k+1}(\tau) \eq \frac{\tau^k}{k!}
\left[ c_{1,0} -C_1^{*} - (N_0-N^{*}) \frac{\tau}{k+1} \right] e^{-\Lambda \tau} \nt
&+& \frac{(\Lambda-1)^2  \gamma\left( k, \Lambda \tau \right)
}{\Lambda^{k+2} (k-1)!} + c_{k,0}\,e^{-\Lambda \tau}  \eeq
Here $c_{k,0}=c_k(0)$ and we use the shorthand notations $N^{*} \equiv  \lambda/(1+\lambda)$ for the equilibrium cluster density and $C_1^{*} \equiv  \lambda^2/(1+\lambda)^2 $ for the equilibrium density of monomers. 

There are again three regimes. In the super-critical regime,  $c_1(\tau ) >0$ for all $\tau$. In this case $\tau
\to \infty$ as $t \to \infty$ and the  final distribution \eqref{final:super} of cluster sizes $C_k$ is
universal and independent on the initial conditions.  In the  sub-critical regime,  $c_1(\tau) \geq 0$ for $\tau \leq \tau_{\rm max}$ and $c_1(\tau) < 0$  otherwise. In the critical regime, $c_1(\tau) \geq 0$ for all $\tau$, and $c_1(\tau_{\rm max})=0 $. For the critical and sub-critical regimes, $\tau \to \tau_{\rm max} < \infty$ as $t \to \infty$ and the distribution $C_k$ is not universal, namely it depends on the initial conditions.

\begin{figure}
\centering
\includegraphics[width=8.5cm]{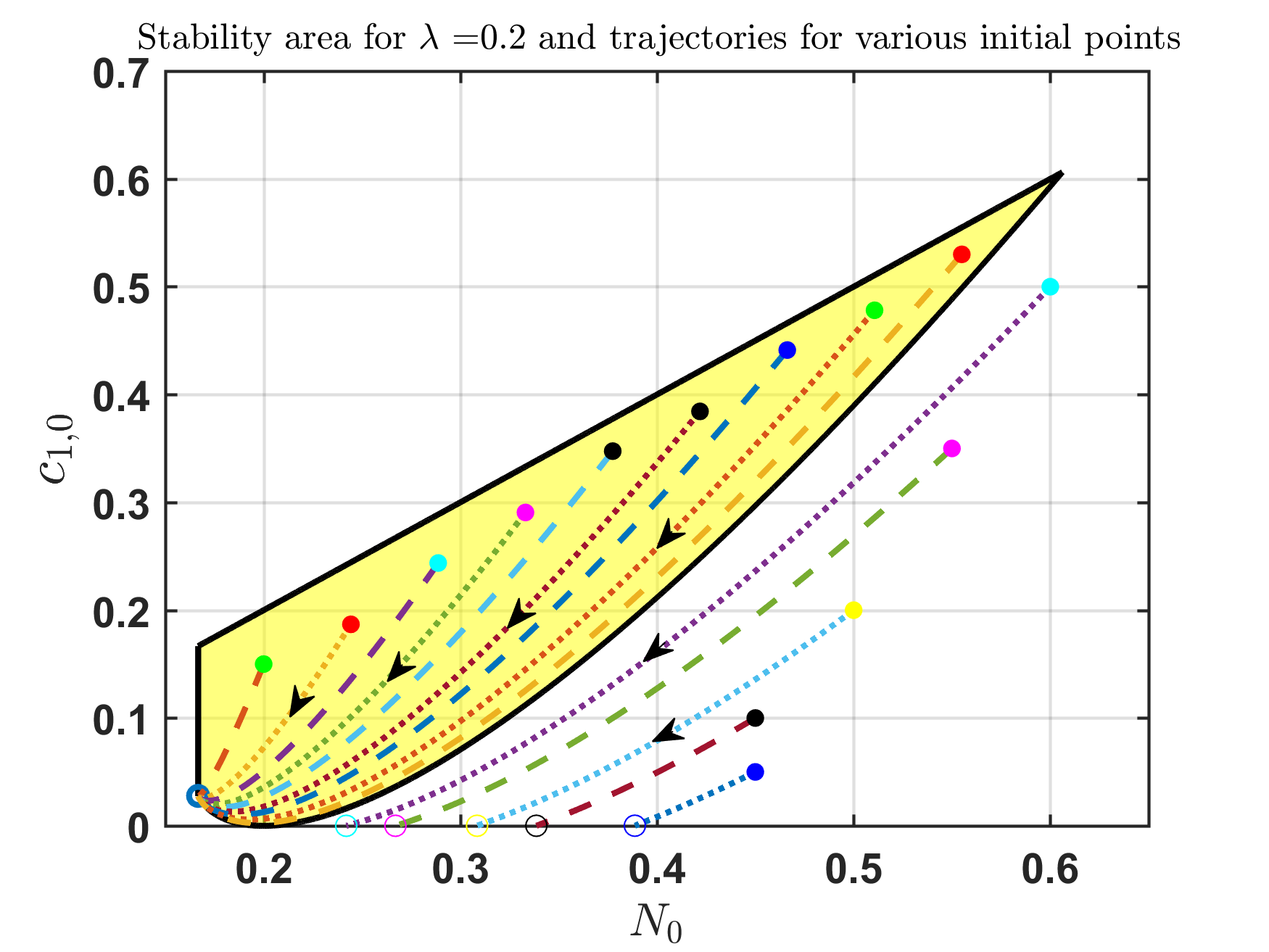} 
\includegraphics[width=8.5cm]{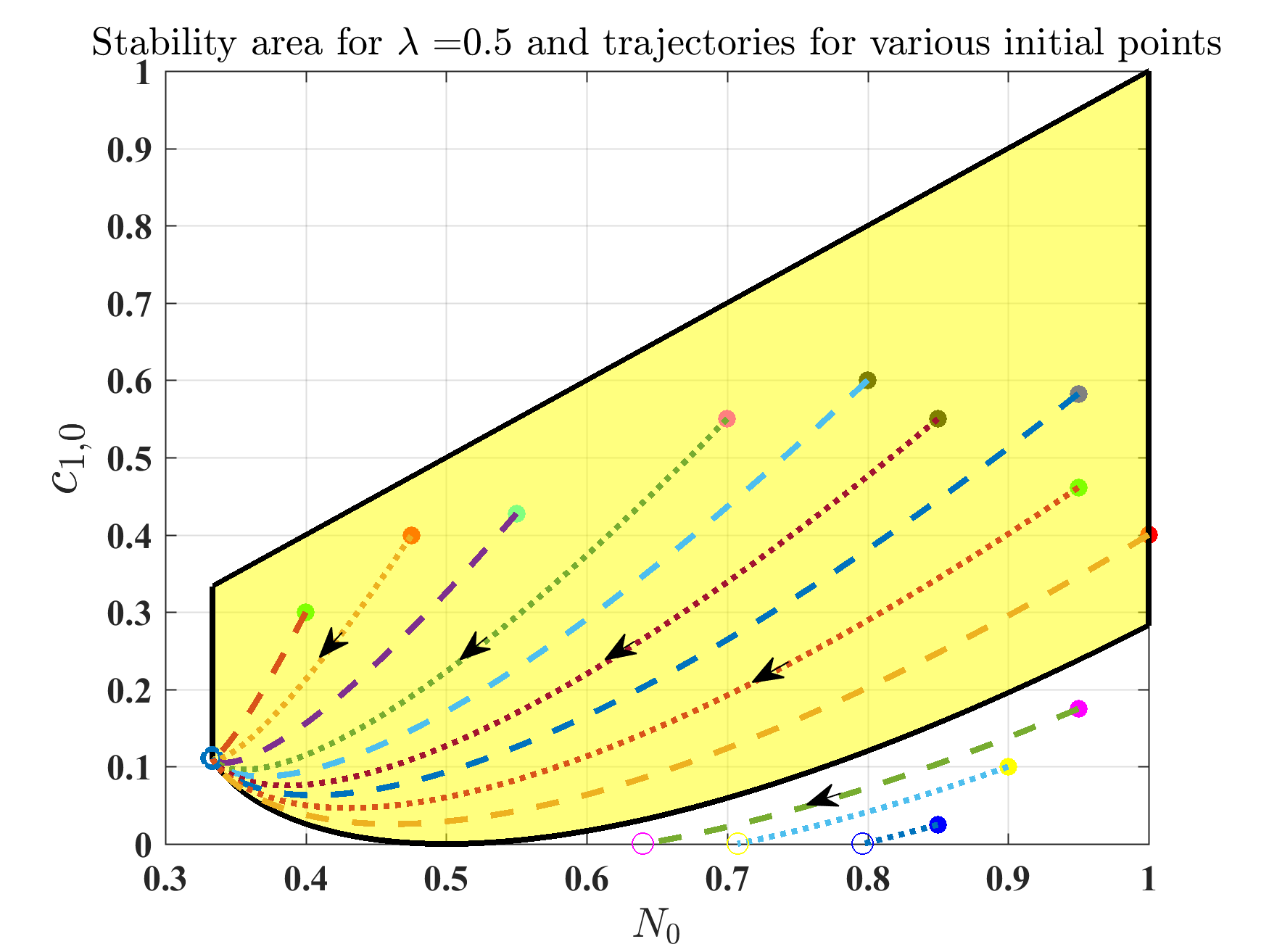}
\caption{Phase trajectories for the total density $N(t)$ and the density of monomers $c_1(t)$ for different initial conditions. The region marked by yellow corresponds to the super-critical regime, where all trajectories terminate at the universal point,
$N^{*}=\lambda/(1+\lambda)$, $C_1^{*}=\lambda^2/(1+\lambda)^2$.  The curved part of there boundary of the yellow domain corresponds to the critical regime. For the initial conditions, located
outside the yellow domain (and satisfying $c_{1,0} \leq N_0$) the sub-critical regime is realized;  the trajectories
terminate in this case on the axis $c_1=0$.}
\label{Fig:domain}
\end{figure}

We emphasize that two parameters, $N_0$ and $c_{1,0}$, play the crucial role in the critical and sub-critical regimes as they demarcate the emergence of these regimes and determine the final modified time $\tau_{\rm max}$. Using \eqref{c1:gen} one can find the domain in the $(N_0, c_{1,0})$ phase plane where $c_1(\tau) >0$ for all $\tau$ \footnote{This may be
done in the same manner as for the case of mono-disperse initial conditions: First, we find $\tau_{\rm max}$
from the condition $dc_1(\tau)/d\tau =0 $ at $\tau = \tau_{\rm max}$,  then the condition of interest reads,
$c_1( \tau_{\rm max}) >0$. It may be recast into the form of Eq.~\eqref{lam_c10N0}}

\be \lb{lam_c10N0} \frac{(1+\lambda)c_{1,0} + N_0 -\lambda }{N_0-N^{*}} > \ln \left(
\frac{N_0-N^{*}}{\lambda N^{*}} \right)  . \ee
(Needless to say, $c_{1,0} \leq N_0$ should be also obeyed).  Within this domain, all trajectories in the $(N, c_1)$ plane terminate at the universal point $(N^{*}, \, C_1^{*})$. If  Eq.~\eqref{lam_c10N0} turns into equality, it determines the critical shattering rate $\lambda_c$ thereby 
yielding the dependence on the initial conditions: $\lambda_c= \lambda_c(N_0, c_{1,0})$.

\begin{figure}
\centering
\includegraphics[width=8.5cm]{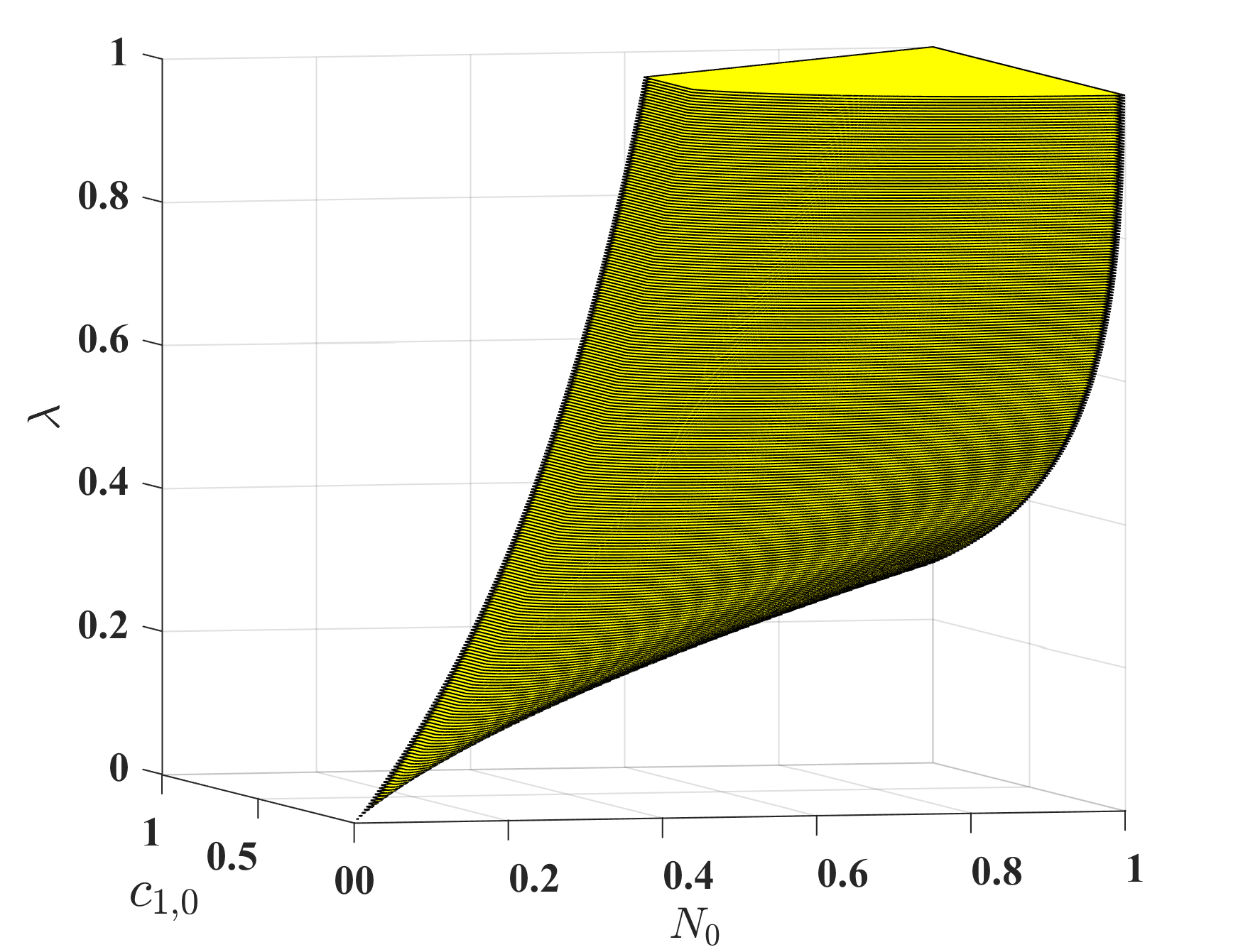} 
\caption{The 3D plot of the critical shattering $\lambda_c$ as the function of the initial conditions $N_0$,
$c_{1,0}$ for the model with constant kinetic rates \eqref{AS:const}. The domain inside the depicted surface corresponds to the super-critical behavior of the system; Fig.~\ref{Fig:domain} provides the cross-sections of this  surface for
particular values of $\lambda$. } \label{Fig3Dlam}
\end{figure}

Outside of the domain given by \eqref{lam_c10N0} the trajectories terminate on the axis $c_1=0$ with $N_{\infty}$ depending on the initial conditions. This is illustrated in Fig.~\ref{Fig:domain}.

Figure \ref{Fig3Dlam}  shows the dependence of the critical shattering $\lambda_c$ on
the initial conditions $N_0$ and $c_{1,0}$. As it may be seen from the Fig.~\ref{Fig3Dlam}, the larger
$\lambda_c$ the larger the domain of the initial conditions corresponding to the super-critical regime, where
the system eventually arrives at the universal steady-state.

\section{Addition and Shattering with Arbitrary Rates}
\label{sec:general}

In the previous  section we studied the addition and shattering processes with mass-independent rates \eqref{AS:const}. A more general class of models is characterized by rates which vary algebraically: $A_k = k^a$ and $S_k=\lambda k^s$. In this section we limit ourselves for the case of $a=s$. We start with $a=s=1$, which admits some analytical treatment.

\subsection{Kinetic rates proportional to the cluster mass}
\label{sec:11}

For the linear dependence of the kinetic rates on the cluster mass, $A_k=k$ and $S_k=\lambda k$, the kinetic
equations read
\begin{subequations}
\begin{align}
&\frac{d c_1}{d\tau}=-(1+\lambda)c_1 - 1 + \lambda M_2
\label{11c1}\\
&\frac{d c_k}{d\tau}=(k-1)c_{k-1}-(1+\lambda)kc_k, \quad k\geq 2
\label{11ck}
\end{align}
\end{subequations}
The second moment $M_2=\sum_{k\geq 1} k^2 c_k$ of the cluster size distribution appears in \eqref{11c1}. Generally the $\alpha^\text{th}$ moment is defined by 

\be \lb{eq:momdef} M_{\alpha}= \sum_{i=1}^{\infty} k^{\alpha} c_k \ee

To determine the monomer density one needs to know $M_2$. The rate equation for the second moment is simple,
\begin{equation}
\label{11M2}
\frac{d M_2}{d \tau} = 2M_2+\lambda(M_2-M_3),
\end{equation}
yet it involves the third moment. The rate equation for the third moment similarly involves the forth moment. This continues ad infinitum leading to (seemingly) unsolvable hierarchy.

\subsubsection{Super-critical regime}

For the super-critical region $\lambda>\lambda_c$ we easily find the equilibrium densities.
Indeed, Eq.~\eqref{11ck} turns into the simple recurrence $(1+\lambda)k C_k = (k-1) C_{k-1}$ admitting an
exact solution giving the equilibrium cluster size distribution
$$
C_k=\frac{C_1}{k (1+\lambda)^{k-1}} 
$$
Using the condition for the mass density, $\sum_{k\geq 1}kC_k=1$, we fix $C_1=\lambda/(1+\lambda)$, which
yields
\begin{equation}
\label{Ck11:steady} C_k = \frac{\lambda}{(1+\lambda)^k}\,\frac{1}{k} =C_k^*
\end{equation}

\subsubsection{Sub-critical and critical regime}

First we consider the mono-disperse initial condition. Applying again the Laplace
transform to \eqref{11ck} yields
\begin{equation}
\label{11ck:rec}
p \widehat{c}_k = (k-1) \widehat{c}_{k-1}-(1+\lambda)k \widehat{c}_k, \quad k\geq 2
\end{equation}
Solving this recurrence  we express all $\widehat{c}_k(p)$ through $\widehat{c}_1(p)$:
\begin{equation}
\label{11ck:sol}
\widehat{c}_k(p) = \frac{\widehat{c}_1(p)}{(1+\lambda)^{k-1}}\,\frac{\Gamma(k)\Gamma(1+\Pi)}{\Gamma(k+\Pi)}
\end{equation}
with $\Pi = 1 + (1+\lambda)^{-1}p$. Applying the Laplace transform for the mass conservation \eqref{mass} we
obtain

\begin{equation}
\label{Lap_mass}
\sum_{k\geq 1} k \widehat{c}_k(p) = \frac{1}{p}.
\end{equation}
Plugging then \eqref{11ck:sol}  into \eqref{Lap_mass} we find the Laplace transform of the monomer density:
\begin{equation}
\label{c1:sol}
\widehat{c}_1 = p^{-1}\,\frac{1}{F[1,2; 1+\Pi; (1+\lambda)^{-1}]}
\end{equation}
where $F[a,b;c;z] = \sum_{n\geq 0}\frac{(a)_n (b)_n}{(c)_n}\,\frac{z^n}{n!}$ is the hypergeometric function.

Near the origin, $p\to 0$~\footnote{The condition $p \to 0$ corresponds to $\tau \to \infty$, that is to the
super-critical case.}, we have $\Pi\to 2$, and using identity $F[1,2;2;z]=(1-z)^{-1}$,
we find that $\widehat{c}_1$ has a simple pole with residue $\lambda/(1+\lambda)$. This is consistent
with the monomer density $c_1(\tau)$ quickly approaching the steady-state value $C_1=\lambda/(1+\lambda)$
which agrees with \eqref{Ck11:steady}.

The precise value of the critical shattering amplitude $\lambda_c$ corresponding to  the mono-disperse
initial condition is hidden in the exact Laplace transform of the monomer
density \eqref{c1:sol}, but difficult to extract. We now show that $\lambda_c>0$ at least for some
initial conditions.

\begin{figure}
\centering
\includegraphics[width=8.5cm]{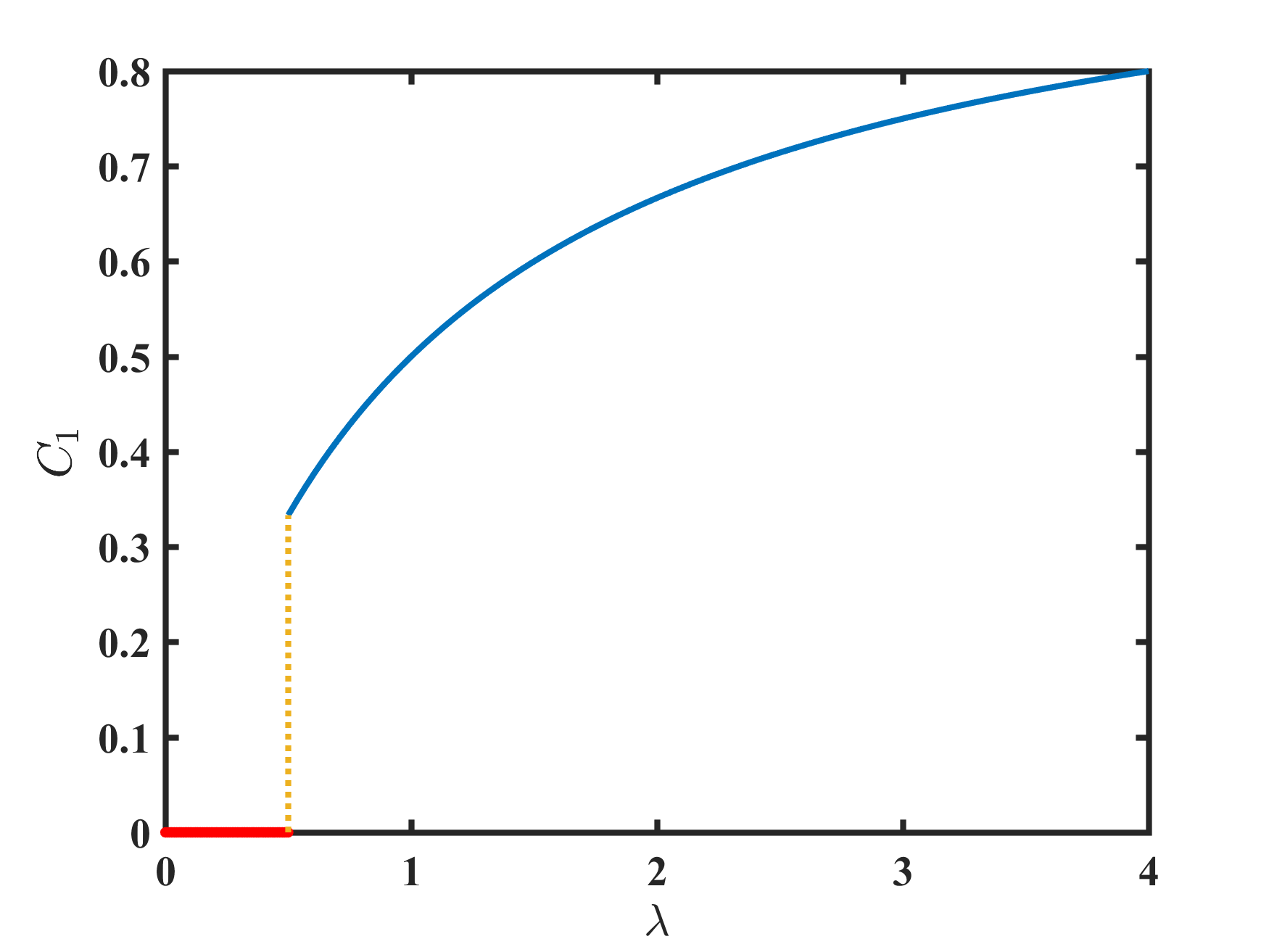}
\caption{For the case of linear kernels, $A_k=k$ and $S_k=\lambda k$, the final density of monomers undergoes
a first order phase transition at $\lambda=\lambda_c=\frac12$. The situation when initially almost all
clusters are dimers is shown. The final density of monomers is given by \eqref{C1:11} in this case. }
\label{Fig:C1_11}
\end{figure}

\begin{figure}
\centering
\includegraphics[width=8.5cm]{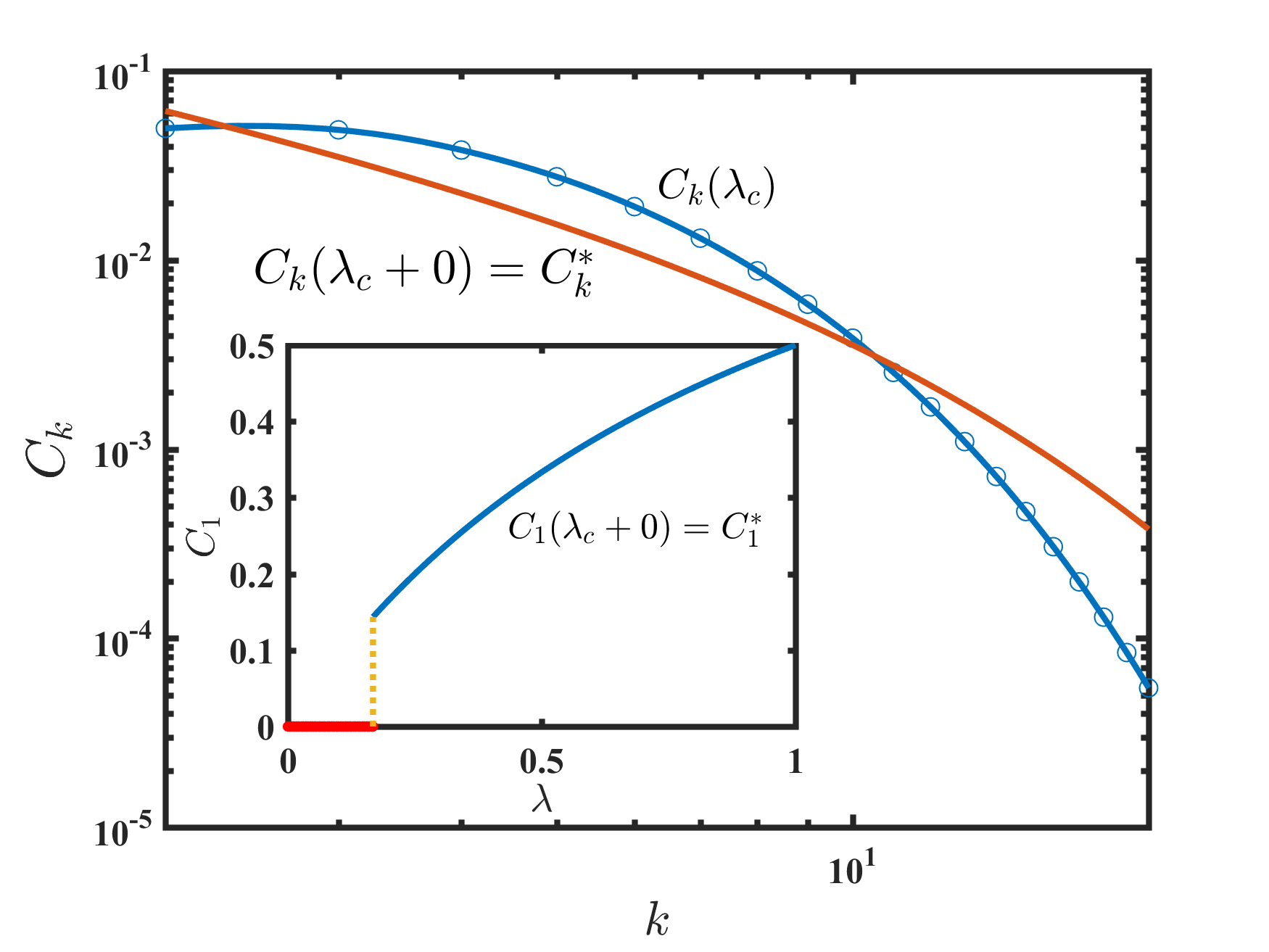} 
\caption{For the case of linear kernels, $A_k=k$ and $S_k=\lambda k$, the discontinuous jump in the cluster
size distribution from $C_k(\lambda_c)$ to $C_k(\lambda_c+0)$ [see Eq.~\eqref{Ck11:steady}] happens for
mono-disperse initial conditions at the critical point $\lambda_c= 0.16773277\ldots$. Inset: The final
density of monomers $C_1(\lambda)$ vanishes in the range $0\leq \lambda\leq \lambda_c$ and  undergoes a first
order transition at $\lambda_c$. In the super-critical region $\lambda>\lambda_c$, $C_1$ is given by
Eq.~\eqref{Ck11:steady}. }
 \label{Fig:CkC1_a1}
\end{figure}

Let us consider a simple case when initially only monomers and dimers present in the system, so that
\be \lb{mondim} c_{1,0}=2N_0 -1; \q  c_{2,0} = 1-N_0; \q M_2(0)= 3-2N_0. \ee
Equation \eqref{11c1} becomes
\begin{equation*}
\frac{d c_1}{d\tau}= - (1+\lambda)c_1 - 1 + \lambda(3-2N_0)
\end{equation*}
when $\tau\ll 1$. Let almost all clusters are dimers, then starting with $c_{1,0}=2N_0 -1\ll 1$,
the monomer density quickly crosses zero if
\begin{equation*}
\lambda(3-2N_0) < 1+(1+\lambda)(2N_0-1)
\end{equation*}
that is, $\lambda<N_0/2(1-N_0)$.  Since $N_0 =c_{1,0}+c_{2,0} \simeq c_{2,0}$ and $M=1=c_{1,0}+2c_{2,0} \simeq 2 c_{2,0}$,
we conclude that $\lambda_c\to \frac{1}{2}$.  Thus in this case (see also Fig.~\ref{Fig:C1_11})
\begin{equation}
\label{C1:11}
C_1 =
\begin{cases}
0                                                & \lambda < \frac{1}{2}\\
 \frac{\lambda}{1+\lambda}        & \lambda > \frac{1}{2}.
 \end{cases}
\end{equation}
For the mono-disperse initial conditions the critical value of $\lambda$ may be found numerically, $\lambda_c
=0.16773277\ldots$. The final concentrations of clusters $C_k(\lambda_c)$, and monomers $C_1(\lambda_c)$,
also found numerically, undergo a first-order phase transition to the values $C_k(\lambda_c+0)$, and monomers
$C_1(\lambda_c+0)$, given by Eq.~\eqref{Ck11:steady}, see Fig.~\ref{Fig:CkC1_a1}.

\begin{figure}
\centering
\includegraphics[width=8.5cm]{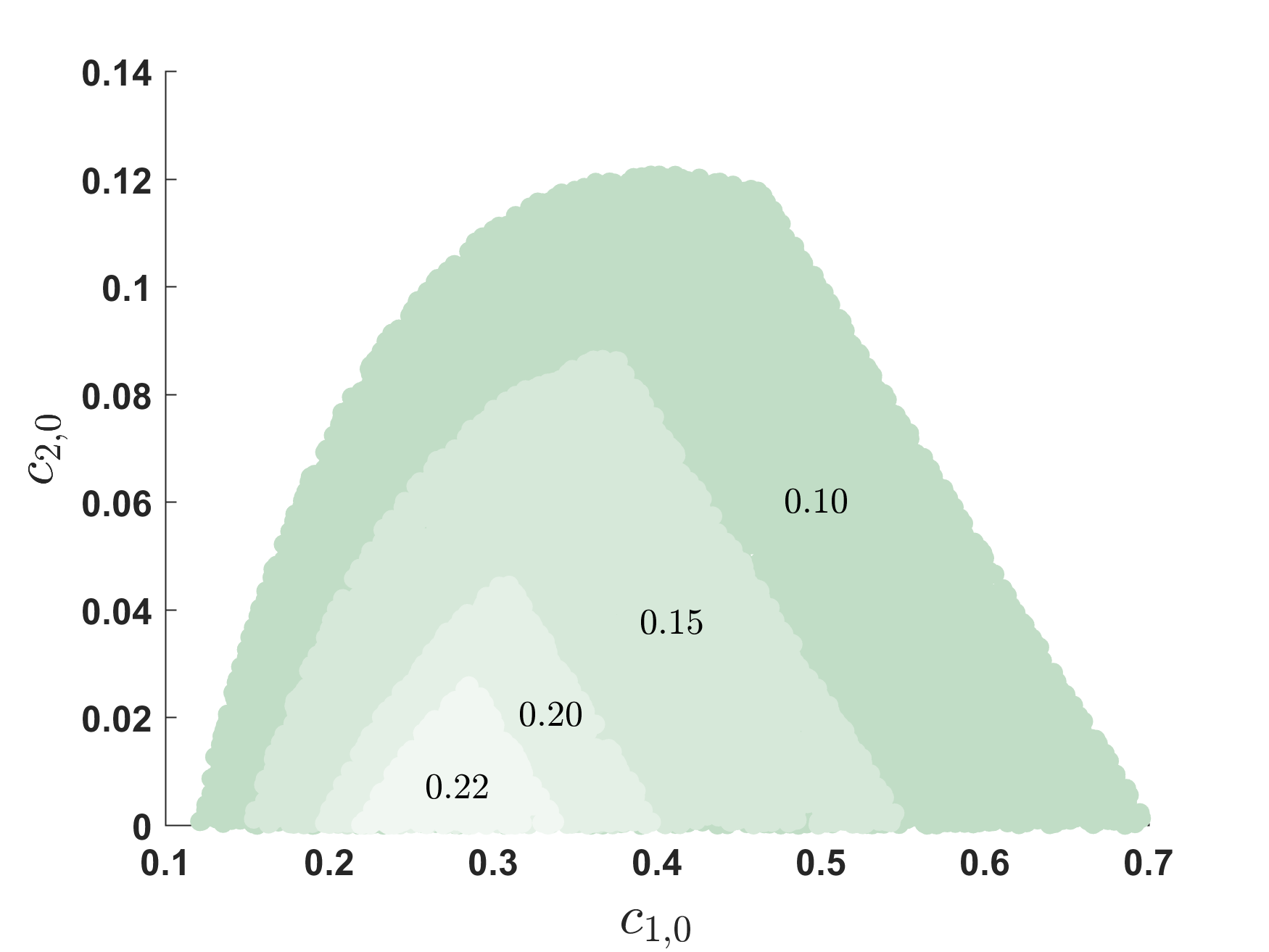} 
\caption{The boundaries of the super-critical region in the phase plane $\left( c_{1,0},\,
c_{2,0} \right)$ for different values of $c_{3,0}$. For each contour, with the value of $c_{3,0}$  indicated
on the plot, the super-critical region lies inside the contour. The  initial densities $c_{k,0}=0$
for $k\geq 5$ and $c_{4,0} =\left(1-c_{1,0}-2c_{2,0}-3c_{3,0} \right)/4$ have been chosen. The
rate kernels are $A_k=k$, $S_k=\lambda k$ and $\lambda=0.16773277\ldots$
(which corresponds to $\lambda_c$ for the mono-disperse initial condition).} 
\label{Fig:lamcc1c2c3}
\end{figure}

The dependence of the evolution regime on the initial conditions is not anymore simple, as in the case of
constant kernels, when only the concentration of monomers and the total number of clusters were important.
For $A_k=k$ and $S_k=\lambda k$ all initial concentrations $c_k(0)$ are determinative.
Fig.~\ref{Fig:lamcc1c2c3} shows the domain in the space of the initial concentrations of the monomers,
$c_{1,0}$, dimers $c_{2,0}$ and trimers, $c_{3,0}$, corresponding to the supercritical regime, when
$c_{4,0}=(1-c_{1,0}-2c_{2,0}-3c_{3,0})/4$ and all other initial concentrations are zero, $c_k(0)=0$ for $k
\geq 5$.

\subsection{Kinetic rates $A_k=k^s$ and $S_k=\lambda k^s$. }

Here we consider a family of models with rates $A_k=k^s$ and $S_k=\lambda k^s$. We assume that exponent does not exceed unity, $s \leq 1$. This is physically motivated (the rates cannot increase faster than mass) and the models with $s>1$ are mathematically ill-defined as instantaneous gelation may occur (it certainly occurs for the pure addition process, see \cite{bk,Laurencot1999}).  

The rate equations read

\begin{subequations}
\begin{align}
&\frac{d c_1}{d\tau}=-(1+\lambda)c_1 - M_s + \lambda M_{1+s}
\label{aac1}\\
&\frac{d c_k}{d\tau}=(k-1)^s c_{k-1}-(1+\lambda)k^s c_k, \quad k\geq 2.
\label{aack}
\end{align}
\end{subequations}
These equations involve the moments $M_s$ and $M_{1+s}$ that satisfy 
\begin{equation*}
\frac{d M_b}{d\tau}= \sum_{k=1}^{\infty} \left[ (k+1)^b  - (1+\lambda)k^{b}\right]k^s c_k - M_s
+ \lambda M_{1+s}
\end{equation*}
with $b=s$ and $b=1+s$. These equations are not closed and for non-integer $s$ they cannot be even written in terms of the moments only. 

Equations \eqref{aac1}--\eqref{aack} still admit some analytical treatment in the super-critical regime ($\lambda>\lambda_c$). 
The equilibrium densities straightforwardly follow from the recursion $k^s (1+\lambda)C_k = (k-1)^s C_{k-1}$ and the mass density $M=1$. One gets

\be \lb{aackst} C_k=\frac{1}{k^s \Lambda^k {\rm Li}_{s-1} (\Lambda^{-1})}  \ee
where $\Lambda =1+\lambda$ as before, and ${\rm Li}_\nu(x)=\sum_{j\geq 1}x^j/j^\nu$ is the polylogarithm function. We have ${\rm Li}_{0}(x)=x/(1-x)$ and thus for $s=1$ we recover the previous result \eqref{Ck11:steady}.

\begin{figure}
\centering
\includegraphics[width=8.5cm]{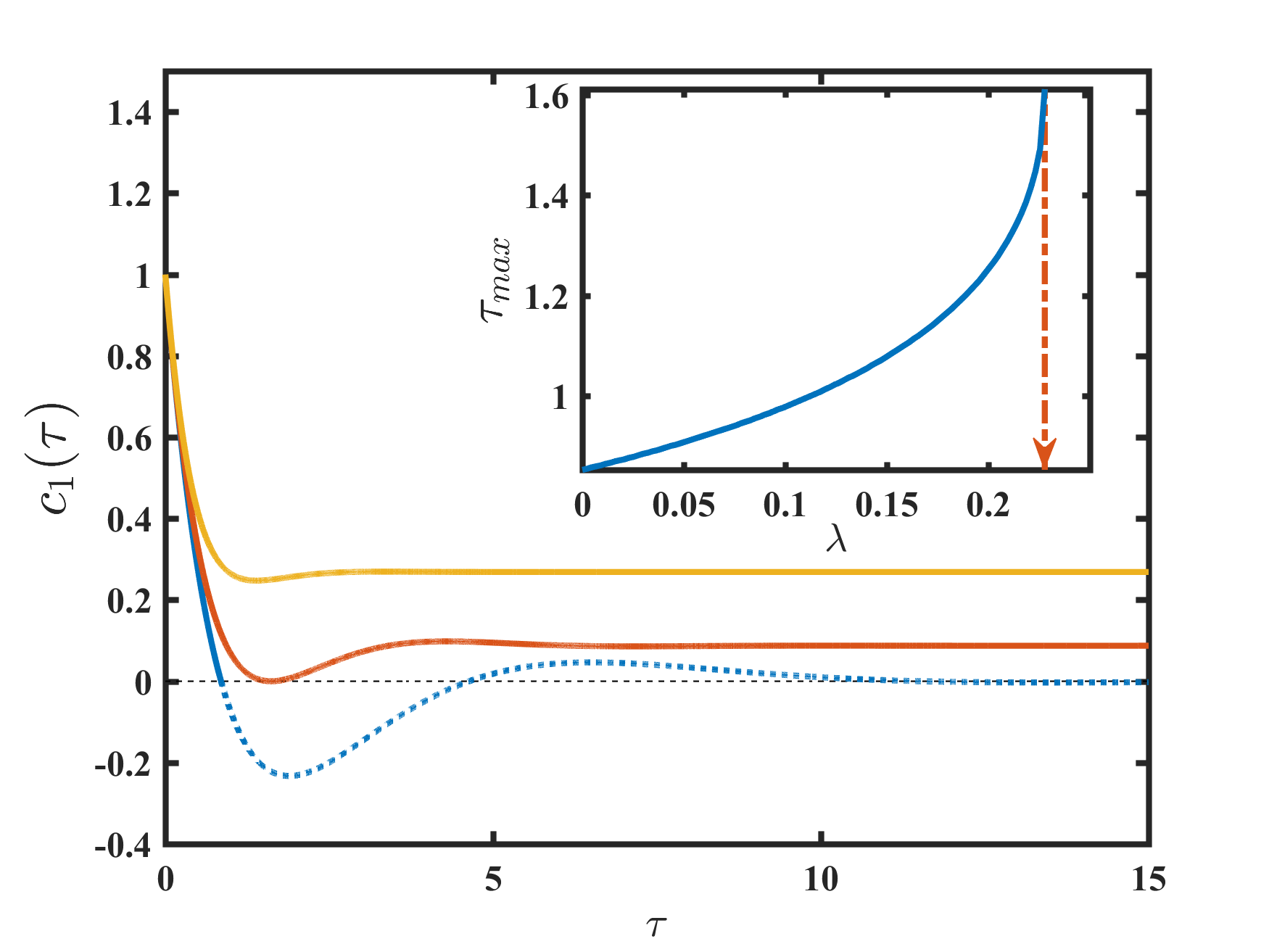}
\caption{Evolution of the monomer density $c_1(\tau)$  for super-critical, critical and sub-critical regimes
for the mono-disperse initial condition. Shown are results for the model with kernels  $A_k=\sqrt{k}$
and $S_k=\lambda \sqrt{k}$. The critical shattering rate in this case is $\lambda_c=0.2277920103...$.} 
\label{Fig:mona05}
\end{figure}

Overall, our simulations show the same qualitative behavior as for $s=0$ and $s=1$. Namely, for any
initial condition there exists a supercritical domain with the final densities given by Eq.~\eqref{aackst}
and $C_1 >0$. This is illustrated in Fig.~\ref{Fig:mona05}, where the evolution of monomer density is
shown for $s=0.5$ for supercritical, critical and sub-critical regimes for the case of mono-disperse initial
conditions.

\begin{figure}
\centering
\includegraphics[width=8.5cm]{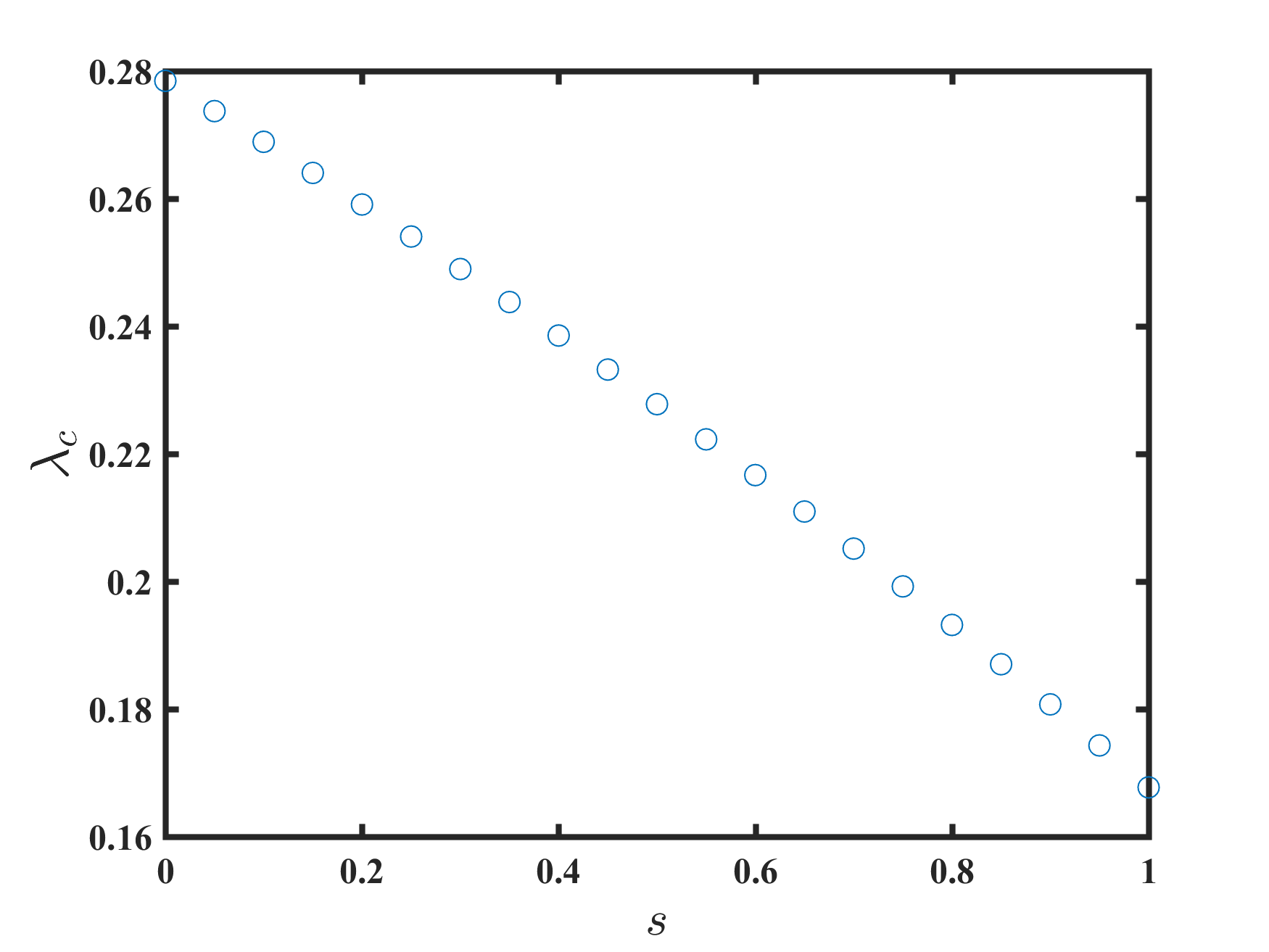}
\caption{The dependence of the critical shattering rate $\lambda_c$ on the exponent $s$ in the case of the mono-disperse initial condition. The critical rate $\lambda_c$ separates the
super-critical evolution regime ($\lambda> \lambda_c$) and sub-critical regime ($\lambda < \lambda_c$). }
\label{Fig:lam_c_of_a}
\end{figure}

The critical shattering rate decreases with increasing $s$, see Fig.~\ref{Fig:lam_c_of_a}, on the interval $0\leq s \leq 1$. The maximum and minimum values of $\lambda_c$ are respectively
$W(1/e)=0.2784645...$ for $s=0$ and $0.1677328...$ for $s=1$.

As in the models with $s=0$ and $s=1$, the final densities of clusters $C_k$ and monomers $C_1$ undergo a jump at the critical point from $C_k(\lambda_c)$ to the super-critical values $C_k(\lambda_c+0)$ given by \eqref{aackst}. As
previously, the final monomer density vanishes for the sub-critical and critical case, i.e. $C_1=0$ for
$\lambda \leq \lambda_c$. The plots for any $0<s<1$ look very similar to Figs.~\ref{Fig:C1_inf}
and \ref{Fig:CkC1_a1}.

\begin{figure}
\centering
\includegraphics[width=8.5cm]{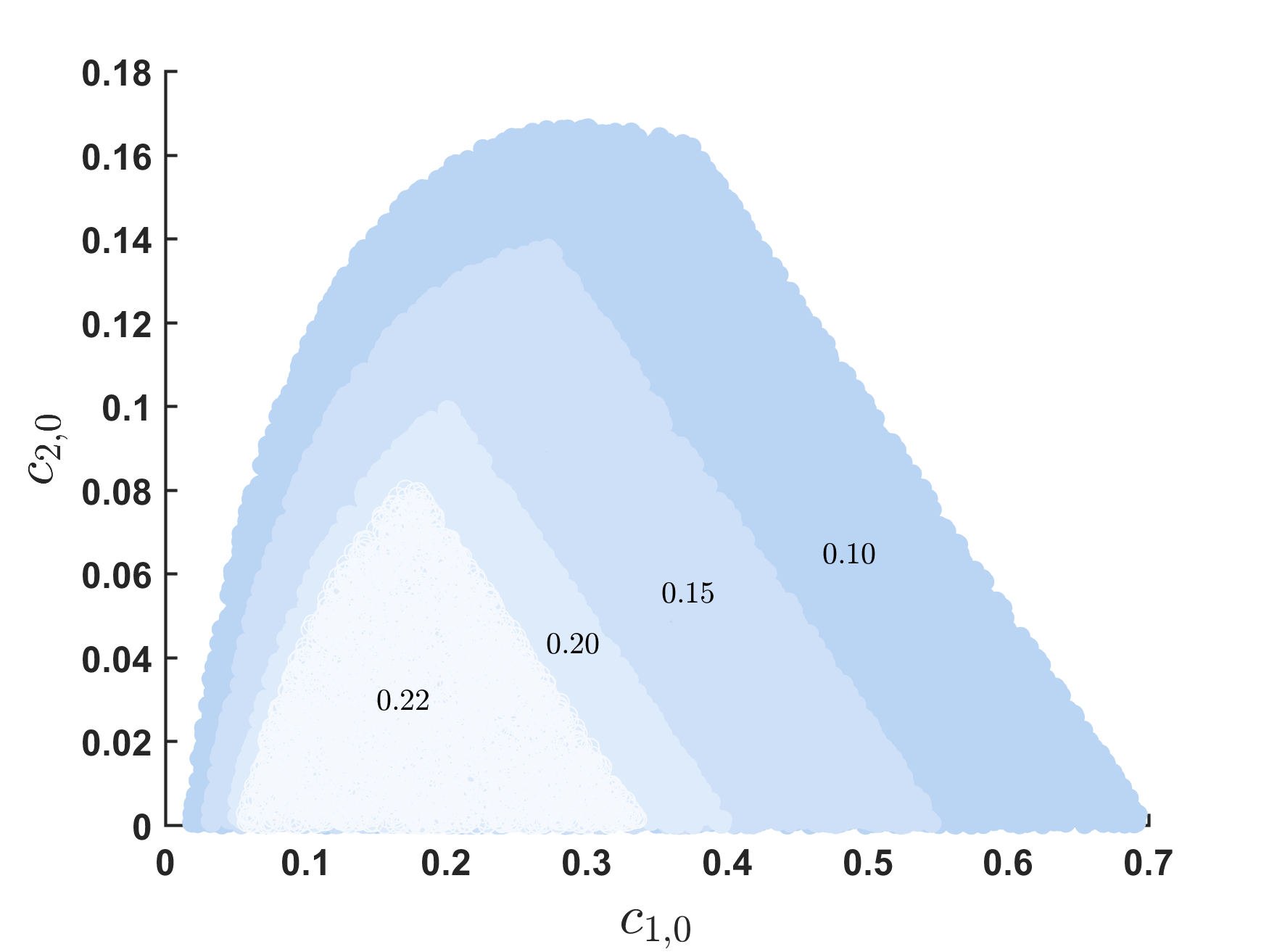} 
\caption{The boundaries of the super-critical region in the phase plane $\left( c_{1,0},
c_{2,0} \right)$ for different values of $c_{3,0}$.  Shown are results for the model with kernels  $A_k=\sqrt{k}$
and $S_k=\lambda \sqrt{k}$ in the case when the initial densities are $c_{k,0}=0$ for $k\geq 5$ and $c_{4,0} =\left(1-c_{1,0}-2c_{2,0}-3c_{3,0} \right)/4$. The shattering rate is $\lambda=0.2277920103 \ldots$ 
(it corresponds to $\lambda_c$ for the mono-disperse initial condition).} \label{Fig:lamca05c1c2}
\end{figure}

The dependence of the critical shattering on the initial conditions is again not simple, since $\lambda_c$
depends on all initial concentrations $c_k(0)$, $k \geq 1$. Accordingly, the super-critical domain depends
for each $\lambda >0$ on all initial concentrations. This is illustrated in Fig.~\ref{Fig:lamca05c1c2}, where
the boundaries of the super-critical domain are shown.

\section{Conclusion}

We investigated a model involving aggregation and fragmentation kinetics where immobile clusters (islands)
interact with mobile monomers. The outcome of the monomer-cluster interactions may be twofold---the monomer
may attach to the  cluster, or the cluster may shatter into monomers. This model mimics the
growth of islands on a surface in epitaxy processes when a collision of an adatom (monomer) with an island
is accompanied by the transmission of the adatom's energy to the island. If the transmitted energy is small,
the adatom joins the island; if the energy is large, the island may fragment into adatoms. We considered a
model when the aggregation rate of a monomer and a cluster of size $k$ depends on $k$ algebraically:  $A_k =k^s$. We assumed that the shattering rate at monomer-cluster collisions is proportional to the addition rate,  $S_k =\lambda A_k$, where $\lambda$ quantifies the shattering intensity. 

In the model with $s=0$ the aggregation and shattering rate do not depend on the island size, $A_k=1$ and
$S_k=\lambda$, so the parameter $\lambda$ characterizes the ratio of shattering and aggregative impacts.
This model admits a comprehensive analytical analysis and demonstrates the most prominent features of
aggregating and shattering systems. We also analyzed the model with $s=1$, and studied numerically the models with 
$0 < s  < 1$. All these models are characterized by three different evolution regimes. In the super-critical regime, $\lambda
>\lambda_c$, the system evolves to a final equilibrium state which is universal, i.e., it does not depend on the
initial conditions. For the critical and sub-critical regimes, $\lambda  \leq \lambda_c$, the evolution terminates at a non-equilibrium {\em jammed} state depending on the initial conditions. In sub-critical regimes,  $\lambda  < \lambda_c$, the evolution to a final jammed state is exponentially fast in time; in the critical regime, $\lambda  = \lambda_c$, the evolution is algebraic. The transition from an equilibrium to a jammed state is a first-order phase transition: The final cluster concentrations $C_k$ undergo a discontinuous jump so that  $C_k(\lambda_c+0) \neq C_k(\lambda_c)$ for $k=1,2, \ldots$. The first-order character of the phase transition is particularly evident in the case of monomers---their density vanishes for $\lambda  \leq \lambda_c$ and remains positive in the super-critical region $\lambda  > \lambda_c$.

A peculiar feature of our system is the dependence on the initial conditions. For the case of constant kernels this dependence  is a very specific one --- only the initial monomer density $c_{1,0}$ and the total density of clusters $N_0$ are important. For each value of $\lambda >0$ there exists a domain in the phase plane $(N_0, c_{1,0})$ that corresponds to the super-critical regime where the system evolves to the universal equilibrium state. For the initial conditions outside this domain the system evolution terminates at jammed states. The size of this domain increases with the increasing $\lambda$.

For the case of general $0 < s  <  1$ the dependence of the evolution regime is more complicated. Although for
all $\lambda >0$ there still exists a domain in the space of initial conditions $\left\{ c_1(0), c_2(0), \ldots
c_k(0), \ldots \right\}$, corresponding to the super-critical behavior, the location of this domain is
determined by all initial densities $c_k(0)$.

In spite of the simplicity of the model, it reflects some prominent features of the processes of the surface films growth. Hence the possibility of the phase transition in such systems, as well as their dependence on the  initial conditions, may be employed for practical manipulations  of the surface films properties. It would be also interesting to study similar models in the context of self-assembly, perhaps generalizing the model to allow a few different types of monomers.

\bibliography{oscillations}

\begin{thebibliography}{22}
\expandafter\ifx\csname natexlab\endcsname\relax\def\natexlab#1{#1}\fi
\expandafter\ifx\csname bibnamefont\endcsname\relax
  \def\bibnamefont#1{#1}\fi
\expandafter\ifx\csname bibfnamefont\endcsname\relax
  \def\bibfnamefont#1{#1}\fi
\expandafter\ifx\csname citenamefont\endcsname\relax
  \def\citenamefont#1{#1}\fi
\expandafter\ifx\csname url\endcsname\relax
  \def\url#1{\texttt{#1}}\fi
\expandafter\ifx\csname urlprefix\endcsname\relax\def\urlprefix{URL }\fi
\providecommand{\bibinfo}[2]{#2}
\providecommand{\eprint}[2][]{\url{#2}}

\bibitem[{\citenamefont{Smoluchowski}(1916)}]{smoluchowski1916drei}
\bibinfo{author}{\bibfnamefont{M.~V.} \bibnamefont{Smoluchowski}},
  \bibinfo{journal}{Z. Phys.} \textbf{\bibinfo{volume}{17}},
  \bibinfo{pages}{557} (\bibinfo{year}{1916}).

\bibitem[{\citenamefont{Krapivsky et~al.}(2010)\citenamefont{Krapivsky, Redner,
  and Ben-Naim}}]{Krapivsky}
\bibinfo{author}{\bibfnamefont{P.~L.} \bibnamefont{Krapivsky}},
  \bibinfo{author}{\bibfnamefont{S.}~\bibnamefont{Redner}}, \bibnamefont{and}
  \bibinfo{author}{\bibfnamefont{E.}~\bibnamefont{Ben-Naim}},
  \emph{\bibinfo{title}{A Kinetic View of Statistical Physics}}
  (\bibinfo{publisher}{Cambridge University Press}, \bibinfo{year}{2010}).

\bibitem[{\citenamefont{Leyvraz}(2003)}]{Leyvraz}
\bibinfo{author}{\bibfnamefont{F.}~\bibnamefont{Leyvraz}},
  \bibinfo{journal}{Physics Reports} \textbf{\bibinfo{volume}{383}},
  \bibinfo{pages}{95} (\bibinfo{year}{2003}).

\bibitem[{\citenamefont{Brilliantov and Krapivsky}(1991)}]{bk}
\bibinfo{author}{\bibfnamefont{N.~V.} \bibnamefont{Brilliantov}}
  \bibnamefont{and} \bibinfo{author}{\bibfnamefont{P.~L.}
  \bibnamefont{Krapivsky}}, \bibinfo{journal}{J. Phys. A}
  \textbf{\bibinfo{volume}{24}}, \bibinfo{pages}{4787} (\bibinfo{year}{1991}).

\bibitem[{\citenamefont{Blackman and Wielding}(1991)}]{Blackman91}
\bibinfo{author}{\bibfnamefont{J.~A.} \bibnamefont{Blackman}} \bibnamefont{and}
  \bibinfo{author}{\bibfnamefont{A.}~\bibnamefont{Wielding}},
  \bibinfo{journal}{EPL} \textbf{\bibinfo{volume}{16}}, \bibinfo{pages}{115}
  (\bibinfo{year}{1991}).

\bibitem[{\citenamefont{Blackman and Marshall}(1994)}]{Blackman94}
\bibinfo{author}{\bibfnamefont{J.~A.} \bibnamefont{Blackman}} \bibnamefont{and}
  \bibinfo{author}{\bibfnamefont{A.}~\bibnamefont{Marshall}},
  \bibinfo{journal}{J. Phys. A} \textbf{\bibinfo{volume}{27}},
  \bibinfo{pages}{725} (\bibinfo{year}{1994}).

\bibitem[{\citenamefont{Privman et~al.}(1999)\citenamefont{Privman, Goia, Park,
  and Matijevic}}]{Privman1999}
\bibinfo{author}{\bibfnamefont{V.}~\bibnamefont{Privman}},
  \bibinfo{author}{\bibfnamefont{D.~V.} \bibnamefont{Goia}},
  \bibinfo{author}{\bibfnamefont{J.}~\bibnamefont{Park}}, \bibnamefont{and}
  \bibinfo{author}{\bibfnamefont{E.}~\bibnamefont{Matijevic}},
  \bibinfo{journal}{J. Colloid Interface Sci.} \textbf{\bibinfo{volume}{213}},
  \bibinfo{pages}{36} (\bibinfo{year}{1999}).

\bibitem[{\citenamefont{Lauren\c{c}ot}(1999)}]{Laurencot1999}
\bibinfo{author}{\bibfnamefont{P.}~\bibnamefont{Lauren\c{c}ot}},
  \bibinfo{journal}{Nonlinearity} \textbf{\bibinfo{volume}{12}},
  \bibinfo{pages}{229} (\bibinfo{year}{1999}).

\bibitem[{\citenamefont{Gorshkov and Privman}(2010)}]{Privman2010}
\bibinfo{author}{\bibfnamefont{V.}~\bibnamefont{Gorshkov}} \bibnamefont{and}
  \bibinfo{author}{\bibfnamefont{V.}~\bibnamefont{Privman}},
  \bibinfo{journal}{Physica E} \textbf{\bibinfo{volume}{43}},
  \bibinfo{pages}{1} (\bibinfo{year}{2010}).

\bibitem[{\citenamefont{Sevonkaev et~al.}(2013)\citenamefont{Sevonkaev,
  Privman, and Goiaa}}]{Privman2013}
\bibinfo{author}{\bibfnamefont{I.}~\bibnamefont{Sevonkaev}},
  \bibinfo{author}{\bibfnamefont{V.}~\bibnamefont{Privman}}, \bibnamefont{and}
  \bibinfo{author}{\bibfnamefont{D.}~\bibnamefont{Goiaa}}, \bibinfo{journal}{J.
  Chem. Phys.} \textbf{\bibinfo{volume}{138}}, \bibinfo{pages}{014703}
  (\bibinfo{year}{2013}).

\bibitem[{\citenamefont{Rothemund et~al.}(2004)\citenamefont{Rothemund,
  Papadakis, and Winfree}}]{RW04}
\bibinfo{author}{\bibfnamefont{P.~W.~K.} \bibnamefont{Rothemund}},
  \bibinfo{author}{\bibfnamefont{N.}~\bibnamefont{Papadakis}},
  \bibnamefont{and} \bibinfo{author}{\bibfnamefont{E.}~\bibnamefont{Winfree}},
  \bibinfo{journal}{PLoS Biology} \textbf{\bibinfo{volume}{2}},
  \bibinfo{pages}{e424} (\bibinfo{year}{2004}).

\bibitem[{\citenamefont{Ariga et~al.}(2008)\citenamefont{Ariga, Hill, Lee,
  Vinu, Charvet, and Acharya}}]{SA08}
\bibinfo{author}{\bibfnamefont{K.}~\bibnamefont{Ariga}},
  \bibinfo{author}{\bibfnamefont{J.~P.} \bibnamefont{Hill}},
  \bibinfo{author}{\bibfnamefont{M.~V.} \bibnamefont{Lee}},
  \bibinfo{author}{\bibfnamefont{A.}~\bibnamefont{Vinu}},
  \bibinfo{author}{\bibfnamefont{R.}~\bibnamefont{Charvet}}, \bibnamefont{and}
  \bibinfo{author}{\bibfnamefont{S.}~\bibnamefont{Acharya}},
  \bibinfo{journal}{Sci. Tech. Adv. Mater.} \textbf{\bibinfo{volume}{9}},
  \bibinfo{pages}{014109} (\bibinfo{year}{2008}).

\bibitem[{\citenamefont{Privman}(2009)}]{Privman2009}
\bibinfo{author}{\bibfnamefont{V.}~\bibnamefont{Privman}},
  \bibinfo{journal}{Ann. New York Acad. Sci.} \textbf{\bibinfo{volume}{1161}},
  \bibinfo{pages}{508 } (\bibinfo{year}{2009}).

\bibitem[{\citenamefont{Demortire et~al.}(2014)\citenamefont{Demortire,
  Snezhko, Sapozhnikov, Becker, Proslier, and Aranson}}]{Igor14}
\bibinfo{author}{\bibfnamefont{A.}~\bibnamefont{Demortire}},
  \bibinfo{author}{\bibfnamefont{A.}~\bibnamefont{Snezhko}},
  \bibinfo{author}{\bibfnamefont{M.~V.} \bibnamefont{Sapozhnikov}},
  \bibinfo{author}{\bibfnamefont{N.}~\bibnamefont{Becker}},
  \bibinfo{author}{\bibfnamefont{T.}~\bibnamefont{Proslier}}, \bibnamefont{and}
  \bibinfo{author}{\bibfnamefont{I.~S.} \bibnamefont{Aranson}},
  \bibinfo{journal}{Nature Communications} \textbf{\bibinfo{volume}{5}},
  \bibinfo{pages}{3117} (\bibinfo{year}{2014}).

\bibitem[{\citenamefont{Evans and Winfree}(2017)}]{Erik17}
\bibinfo{author}{\bibfnamefont{C.~G.} \bibnamefont{Evans}} \bibnamefont{and}
  \bibinfo{author}{\bibfnamefont{E.}~\bibnamefont{Winfree}},
  \bibinfo{journal}{Chem. Soc. Rev.} \textbf{\bibinfo{volume}{46}},
  \bibinfo{pages}{3808} (\bibinfo{year}{2017}).

\bibitem[{\citenamefont{Bartelt and Evans}(1992)}]{Evans_PRB}
\bibinfo{author}{\bibfnamefont{M.~C.} \bibnamefont{Bartelt}} \bibnamefont{and}
  \bibinfo{author}{\bibfnamefont{J.~W.} \bibnamefont{Evans}},
  \bibinfo{journal}{Phys. Rev. B} \textbf{\bibinfo{volume}{46}},
  \bibinfo{pages}{12675} (\bibinfo{year}{1992}).

\bibitem[{\citenamefont{Kallabis et~al.}(1998)\citenamefont{Kallabis,
  Krapivsky, and Wolf}}]{Wolf}
\bibinfo{author}{\bibfnamefont{H.}~\bibnamefont{Kallabis}},
  \bibinfo{author}{\bibfnamefont{P.~L.} \bibnamefont{Krapivsky}},
  \bibnamefont{and} \bibinfo{author}{\bibfnamefont{D.~E.} \bibnamefont{Wolf}},
  \bibinfo{journal}{EPJB} \textbf{\bibinfo{volume}{5}}, \bibinfo{pages}{801}
  (\bibinfo{year}{1998}).

\bibitem[{\citenamefont{Pimpinelli and Villain}(1998)}]{MBE}
\bibinfo{author}{\bibfnamefont{A.}~\bibnamefont{Pimpinelli}} \bibnamefont{and}
  \bibinfo{author}{\bibfnamefont{J.}~\bibnamefont{Villain}},
  \emph{\bibinfo{title}{Physics of Crystal Growth}}
  (\bibinfo{publisher}{Cambridge University Press}, \bibinfo{year}{1998}).

\bibitem[{\citenamefont{Zinke-Allmang}(1999)}]{Zinke_Allmang1999}
\bibinfo{author}{\bibfnamefont{M.}~\bibnamefont{Zinke-Allmang}},
  \bibinfo{journal}{Thin Solid Films} \textbf{\bibinfo{volume}{346}},
  \bibinfo{pages}{1} (\bibinfo{year}{1999}).

\bibitem[{\citenamefont{Krapivsky et~al.}(1999)\citenamefont{Krapivsky, Mendes,
  and Redner}}]{Krapivsky_PRB1999}
\bibinfo{author}{\bibfnamefont{P.~L.} \bibnamefont{Krapivsky}},
  \bibinfo{author}{\bibfnamefont{J.~F.~F.} \bibnamefont{Mendes}},
  \bibnamefont{and} \bibinfo{author}{\bibfnamefont{S.}~\bibnamefont{Redner}},
  \bibinfo{journal}{Phys. Rev. B} \textbf{\bibinfo{volume}{59}},
  \bibinfo{pages}{15950} (\bibinfo{year}{1999}).

\bibitem[{\citenamefont{Amar et~al.}(2001)\citenamefont{Amar, Popescu, and
  Family}}]{Family2001}
\bibinfo{author}{\bibfnamefont{J.~G.} \bibnamefont{Amar}},
  \bibinfo{author}{\bibfnamefont{M.~N.} \bibnamefont{Popescu}},
  \bibnamefont{and} \bibinfo{author}{\bibfnamefont{F.}~\bibnamefont{Family}},
  \bibinfo{journal}{Phys. Rev. Lett.} \textbf{\bibinfo{volume}{86}},
  \bibinfo{pages}{3092 – 3096} (\bibinfo{year}{2001}).

\bibitem[{\citenamefont{Popescu et~al.}(2001)\citenamefont{Popescu, Amar, and
  Family}}]{Family_PRB2001}
\bibinfo{author}{\bibfnamefont{M.~N.} \bibnamefont{Popescu}},
  \bibinfo{author}{\bibfnamefont{J.~G.} \bibnamefont{Amar}}, \bibnamefont{and}
  \bibinfo{author}{\bibfnamefont{F.}~\bibnamefont{Family}},
  \bibinfo{journal}{Phys. Rev. B} \textbf{\bibinfo{volume}{64}},
  \bibinfo{pages}{205404} (\bibinfo{year}{2001}).

\end{thebibliography}
\end{document}